\documentclass[twocolumn]{aastex62}

\usepackage[caption=false]{subfig}
\usepackage{amsmath,amstext}
\usepackage[T1]{fontenc}
\usepackage[normalem]{ulem}
\usepackage{threeparttable}
\usepackage{url}
\usepackage{xspace}

\usepackage{color}

\DeclareMathAlphabet{\mathsc}{OT1}{cmr}{m}{sc}
\def\testbx{bx}%
\DeclareRobustCommand{\ion}[2]{%
\relax\ifmmode
\ifx\testbx\f@series
{\mathbf{#1\,\mathsc{#2}}}\else
{\mathrm{#1\,\mathsc{#2}}}\fi
\else\textup{#1\,{\mdseries\textsc{#2}}}%
\fi}

\newcommand\blfootnote[1]{%
  \begingroup
  \renewcommand\thefootnote{}\footnote{#1}%
  \addtocounter{footnote}{-1}%
  \endgroup
}
\def\ben{\begin{enumerate}}
\def\een{\end{enumerate}}
\def\bi{\begin{itemize}}
\def\ei{\end{itemize}}
\def\be{\begin{equation}}
\def\ee{\end{equation}}
\def\bea{\begin{eqnarray}}
\def\eea{\end{eqnarray}}

\def\nhi{N_{\ion{H}{i}}}
\def\planck{\textit{Planck}\xspace}
\def\herschel{\textit{Herschel}\xspace}

\def\fsky{$18.3$}
\def\nlb{64\xspace}

\received{May 6, 2019}
\revised{July 29, 2019}
\accepted{August 14, 2019}
\submitjournal{ApJ}

\shorttitle{Large-scale CIB maps from Planck}
\shortauthors{Lenz et al.}

\begin{document}

\title{Large-scale Maps of the Cosmic Infrared Background from \planck}

\author[0000-0001-5820-475X]{Daniel Lenz}
\affiliation{Jet Propulsion Laboratory, California Institute of Technology, Pasadena, CA 91109, USA}
\affiliation{California Institute of Technology, Pasadena, CA, 91125 USA}

\author[0000-0001-7432-2932]{Olivier Dor\'e}
\affiliation{Jet Propulsion Laboratory, California Institute of Technology, Pasadena, CA 91109, USA}
\affiliation{California Institute of Technology, Pasadena, CA, 91125 USA}

\author[0000-0003-1492-2519]{Guilaine Lagache}
\affiliation{Aix Marseille Univ, CNRS, CNES, LAM, Marseille, France}

\email{mail@daniellenz.org}


\begin{abstract}
The cosmic infrared background (CIB) is a powerful probe of large-scale structure across a very large redshift range, and consists of unresolved redshifted infrared emission from dusty galaxies. It can be used to study the astrophysics of galaxies, the star formation history of the universe, and the connection between dark and luminous matter. It can furthermore be used as a tracer of the large-scale structure and thus assist in de-lensing of the cosmic microwave background. The major difficulty in its use lies in obtaining accurate and unbiased large-scale CIB images that are cleaned of the contamination by Galactic dust. We used data on neutral atomic hydrogen from the recently released HI4PI Survey to create template maps of Galactic dust, allowing us to remove this component from the \planck intensity maps from 353 to 857\,GHz for approximately 25\% of the sky. This allows us to constrain the CIB power spectrum down to $\ell\gtrsim 70$. We present these CIB maps and the various processing and validation steps that we have performed to ensure their quality, as well as a comparison with previous studies. All our data products are made publicly available \footnote{\url{https://doi.org/10.7910/DVN/8A1SR3}}, thereby enabling the community to investigate a wide range of questions related to the universe's large-scale structure.
\end{abstract}



\section{Introduction}
\label{sect:intro}

\blfootnote{\copyright 2019 California Institute of Technology.\\ U.S. Government sponsorship acknowledged}

The cosmic infrared background (CIB) consists of the integrated emission from unresolved dusty star forming galaxies \citep{puget1996, gispert2000, lagache2005, dole2006}. This emission stems from dust grains bound to these galaxies and heated by the ultraviolet emission from young stars. As a result, most of the CIB emission originates from the peak of the star formation epoch at $z=1-2$ and originates in galaxies living in dark matter (DM) halos with masses of $10^{11}$ to $10^{13}\, M_{\odot}$ \citep{bethermin2012, schmidt2015}. The CIB is an excellent tool with which to study the cosmic star formation history and the connection between dark and luminous matter.

Historically, the monopole of the CIB was first detected through measurements with the FIRAS and DIRBE instruments aboard the \textit{COBE} satellite \citep{puget1996, fixsen1998}. It took another decade to also detect the anisotropies of the CIB \citep[e.g. in \textit{IRAS} data,][]{miville-deschenes2002}, which enabled the first measurement of the auto power spectrum and constraints on the bias of the CIB sources \citep{lagache2007}. The latest milestone in understanding the CIB was a result of the \planck and \herschel missions \citep{pilbratt2010, planck2018_i}. The combination of high resolution, high sensitivity, broad frequency coverage, and large area allowed for the extraction of unprecedented maps, and for constraints on a wide range of models \citep{planck2011_xviii, planck2014_xxx, serra2014, planck2016_xlviii, mak2017, viero2018}.


On top of being an excellent probe of star formation, the CIB is also a good tracer of the large-scale structures out to high redshift, and thus a good tracer of the lensing potential that affects the cosmic microwave background (CMB) \citep{planck2014_xviii}. As a consequence, the CIB can be used to predict the lensing potential and to de-lens the CMB temperature map as demonstrated, for example, on \planck data \citep{larsen2016}. \citet{manzotti2017} used CIB data from \herschel to de-lens the CMB B-mode map from the South Pole Telescope \citep{carlstrom2011}, thereby detecting for the first time a statistically significant de-lensing effect on the CMB B-mode. Forecasts conducted by \citet{manzotti2018} show that for next-generation CMB Stage-3 and Stage-4 experiments, the internal reconstruction of the lensing potential will be the dominant method to de-lens the CMB, but ancillary data will remain very helpful and serve as a very important and independent systematic check. Most notably, the CIB covers the lensing potential out to much higher redshift than what can be covered in current large-area galaxy surveys, albeit with much lower resolution. Lastly, \citet{planck2018_viii} combined the \planck lensing reconstruction at large scales, which was high signal-to-noise, with CIB information from the \planck generalized needlet internal linear combination (GNILC) maps \citep{planck2016_xlviii} at small scales to construct an optimal map of the lensing potential that combines the advantages of both tracers.

Large-area maps of the CIB are maps of a highly biased tracer of dark matter at $z=1-2$. It is thus a promising avenue to constrain the scale-dependent bias imprinted by primordial non-Gaussianity \citep[e.g.][and references therein]{Dalal:2007cu,deputter2017}. \citet{tucci2016} use Fisher forecasts to demonstrate that, even in the presence of Galactic dust residuals, an uncertainty of $\sigma(f_{\rm NL})$ of approximately 3.5 can be obtained for sky fractions between 0.2 and 0.6. This result would be competitive with CMB bispectrum-based measurements \citep{planck2016_xvii} and would rely on different physical scales. Current CIB-based measurements of this effect are limited by the available area of the CIB maps with low enough residual galactic dust. Our work aims at directly addressing this limitation.

In the present work, we extend the \ion{H}{i} column density-based approach to be more general, less subjective to human decisions, and to cover a larger fraction of the sky. At the same time, we now also have access to newer data, both from \planck for the far-infrared (FIR) data, and from the HI4PI Survey \citep{hi4pi2016} for the \ion{H}{i} data. We present the underlying data sets in Section \ref{sect:data}, our methodology in Section \ref{sect:methods}, the estimate of the power spectrum  in Section \ref{sect:powerspectra}, the map-based results in Section \ref{sect:map_results}, and the results for the power spectra in Section \ref{sect:powerspectra_results}. We present the validation of the results in Section \ref{sect:validation}, and finally conclude in Section \ref{sect:conclusions}.

\section{Data and Preprocessing}
\label{sect:data}

We briefly describe the individual data products used throughout this study and the preprocessing that was performed on the publicly available data.

\subsection{\ion{H}{i} data}
\label{sect:data:hi4pi}

The \ion{H}{i} data are based on the recently published HI4PI Survey \citep{hi4pi2016}. This survey merges data from the Effelsberg--Bonn \ion{H}{i} Survey \citep[EBHIS,][]{winkel2010, kerp2011, winkel2016a} and the Galactic All-Sky Survey \citep[GASS,][]{mcclure-griffiths2009, kalberla2010, kalberla2015} to create a full-sky database of Galactic atomic neutral hydrogen.

Compared to its predecessor, the Leiden/Argentine/Bonn Survey \citep[LAB,][]{kalberla2005}, it offers full spatial sampling (instead of beam-by-beam sampling), higher angular resolution ($16.1'$ instead of $35'$), and higher sensitivity.

Our preprocessing of the HI4PI data is described below. This data set is publicly available on CDS.\footnote{\url{http://cdsarc.u-strasbg.fr/viz-bin/qcat?J/A+A/594/A116}}

\begin{figure}
	\includegraphics[width=\columnwidth]{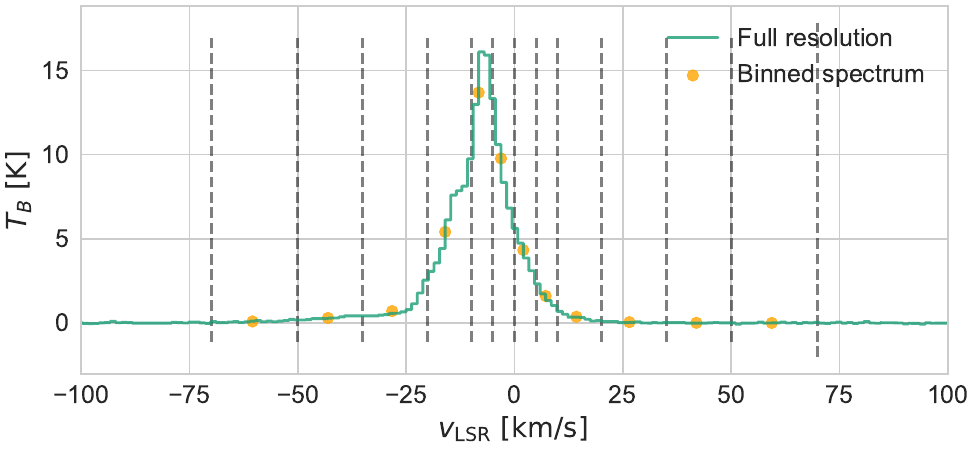}
	\caption{Example of a spectrum at full resolution and of the binning scheme applied for this work. We exclude emission with $|v_{\rm LSR}| > 90\,\rm km/s$ because of its very low dust-to-gas ratio.}
	\label{fig:example_spectra}
\end{figure}

\begin{enumerate}
	\item We merge the individual smaller cubes (each containing data for one HEALPix pixel at Nside 4) into one large HDF5 table.
	\item The Magellanic System, especially the Magellanic Stream, is a major contaminant due to its low dust content despite high \ion{H}{i} column densities. We follow the procedure detailed in A.2 of \citet{planck2014_xxx} to mask this emission in the 3D cube. We use the Milky Way model described in \citet{kalberla2008} and mask all emission with model brightness temperatures less than 60\;mK and Magellanic coordinates $240^{\circ} < \lambda < 30^{\circ}$ and $|\beta| < 10^{\circ}$.

	\item Instead of working with the full spectral resolution of the HI4PI Survey, we bin the data along the spectral axis. We choose a non-uniform binning to capture the highly complex emission close to the velocity of the local standard-of-rest $v_{\rm LSR} = 0$ while using large bins at the higher velocities that contain little emission. This binning scheme is illustrated in Figure \ref{fig:example_spectra}. We also exclude emission with $|v_{\rm LSR}| > 90\,\rm km/s$ because of its very low dust-to-gas ratio \citep{planck2011_xxiv, lenz2016}. This binning scheme is applied to the full sky, and does not differ from sightline to sightline.

\end{enumerate}

\subsection{\planck data}
\label{sect:data:planck}

For the FIR data, we used the Public Release 3 (PR3) data from the \planck satellite. We work with the high-frequency data at 353, 545, and 857\;GHz. For the highest frequencies, the dust emission from the Milky Way and from the CIB dominate over other processes such as CMB, synchrotron, or free-free emission. For the lower frequencies, the \planck bands have a strong CMB component. For consistency, we remove the CMB from all frequencies.

At 217\;GHz, we found the strong CMB contribution to be a limiting factor in the analysis. The resulting CIB map at this frequency depends strongly on the exact details of the CMB correction, and is consequently not presented here.

Our full preprocessing pipeline of the \planck data is described below. All the data products were obtained from the \planck Legacy Archive\footnote{\url{http://pla.esac.esa.int/pla/}}. For most of the analysis, we used the \texttt{healpy}\footnote{\url{https://healpy.readthedocs.io/en/latest/}} implementation \citep{zonca2019} of the HEALPix pixelization \citep{gorski2005}. We refer to the $353-857\;\rm GHz$ data as intensity maps. Some of the operations listed below require converting the maps back and forth from $K_{\rm CMB}$ to $\rm MJy/sr$. This is done by using the conversion factors given in \citet[][Table 3]{planck2016_x}.

\begin{enumerate}
	\item We start with the \planck intensity map in temperature at frequency $\nu$, at full resolution, and HEALPix Nside of 2048 (e.g. \path{HFI_SkyMap_545_2048_R3.00_full-oddring.fits}, using the \texttt{I\_STOKES} data and \texttt{II\_COV} for the covariance).
	
	\item We convert the map from nested ordering to ring ordering.
	\item We subtract from the intensity maps the CIB monopole that was added by hand to the \planck maps at 353, 545, and 857 GHz, following the model of \citet{bethermin2012}. The values are given in \citet[][Table 12]{planck2018_iii}.
	\item Given the beam window function in the \textit{Planck} Reduced Instrument Model (RIMO), we convolve the maps to the window function of the SMICA CMB map. This means converting the map $a_{\ell m}$, multiplying these by $B^{\rm kernel}_{\ell}=B^{\rm\;SMICA}_{\ell}/B^{I_{\nu}}_{\ell}$, and converting back from $a_{\ell m}$ to map.
	\item With the intensity maps now at the same 5' resolution as the CMB map, we subtract the SMICA CMB map from the intensity map. The maps need to be in units of $K_{\rm CMB}$ for this.
	\item Lastly, we downgrade the resolution of the HEALPix grid from an Nside of 2048 to an Nside of 1024, which is sufficient to fully sample the data at this resolution. 
	We only conduct the generation of the foreground dust model at Nside 1024. This model is subsequently upsampled to Nside 2048 for all following analyses. The final published CIB maps are also distributed at Nside 2048.
\end{enumerate}

We note that the beams are slightly asymmetric, which will introduce a 1-2\% error at $\ell\approx 1500$, which we consider as negligible here. This effect has been investigated in Section 4.3 of \citet{planck2018_iv}.

Furthermore, the subtraction of the SMICA CMB map could be biased due to the contamination of that CMB map with residual CIB and Galactic dust. An illustration of the order of magnitude of these effects is given in \citet[][Figure E.6]{planck2015_ix}, showing that the contamination is very minor, in particular for the higher frequencies.

To avoid noise bias when measuring angular power spectra, we are measuring cross spectra between ring maps. The so-called odd- and even-ring maps are generated using only the first or the second half of each pointing period. These maps were used by the \planck team to test internal consistency and to characterize the noise by taking the difference of these half-ring maps. The periods are typically 20 minutes long, and there are two half-ring maps at each frequency. Each of the two half-ring maps at each frequency is processed according to the steps outlined above. We also ran the entire analysis on half-mission maps rather than half-ring maps, and found no significant difference.

\subsection{The \planck lensing map}
\label{sect:data:lensing}

In order to quantify the effectiveness of our Galactic dust cleaning procedure, we will use the publicly available \planck lensing convergence map \citep{planck2018_viii}, which is known to be strongly correlated with the CIB \citep{planck2014_xviii}. To cross-correlate this lensing map with our CIB map, we generate a lensing map at an Nside of 2048, based on the publicly available lensing $a_{\ell m}$ and lensing mask. We note that this map is already corrected for the response function, including the beam function and the pixel window function.

In addition, the lensing angular power spectra are often expressed in terms of the lensing potential $\phi$ and not the lensing convergence $\kappa$. The two are related through
\begin{equation}
	\kappa_{\ell m} = \frac{1}{2}\;\ell\; (\ell + 1)\;\phi_{\ell m} \,.
\end{equation}
In this paper, we express all lensing-related quantities using the lensing convergence $\kappa$.

\subsection{Masks}
\label{sect:data:masks}

To compute an accurate and reliable angular power spectrum of CIB anisotropies, we need to mask different components that would otherwise bias the analysis. Figure \ref{fig:masks} shows the final mask that is used for our analysis. There are several components that contribute to this effective mask.

First of all, we limit our analysis to the low \ion{H}{i} column densities. Based on our experience in \citet{planck2014_xxx} and \citet{lenz2017a}, we choose $\nhi=2.5\times 10^{20}\,\rm cm^{-2}$ as the threshold for our baseline results shown here. In addition to that, we also use the \textit{Planck} 20\% Galactic plane mask, so the final mask is the intersection of the two. The complexity of the interstellar medium (ISM) close to the Galactic disk makes the CIB measurements in this region almost impossible. We further explore different cuts in \ion{H}{i} column density in Section \ref{sect:validation:nhi_thresh}, and publish maps with different sky fractions and thus different levels of Galactic dust residuals.

Second, we employ the public \planck masks of extragalactic point sources, which are limited to a signal-to-noise ratio of 5. For further reference and for the specific flux limits, see \citep{planck2016_xxvi}. These masks are slightly different for each frequency, with the highest, 857\,GHz channel containing the most point sources. Additionally, we intersect them with the effective mask of the SMICA CMB map.

Third, we find that a major source of contamination is molecular gas at high Galactic latitudes, often associated with molecular intermediate-velocity clouds \citep[MIVCs,][]{magnani2010, roehser2016}. On top of that, the linear correlation between \ion{H}{i} column density and FIR dust emission only holds in the absence of CO-dark molecular gas, which cannot be observed directly \citep{planck2011_xix}. A census of these objects was performed in the work of \citet{roehser2016}, and we use a HEALPix mask of all these sources (T. Roehser, private communication). With the $N_\ion{H}{i}$ threshold already in place, this mask only excludes an additional $21\;\rm deg^2$.

Lastly, we mask residual Galactic dust emission in our final CIB maps. In most cases, this residual emission results from dust that is associated not with \ion{H}{i}, but instead with CO-bright or CO-dark molecular gas. For the column densities investigated here, the influence of optically thick \ion{H}{i} is negligible \citep[e.g.][]{lee2015}. We provide further details on this procedure in Section\,\ref{sect:methods}.

\subsubsection{Mask apodization}

\begin{figure}
	\includegraphics[width=\columnwidth]{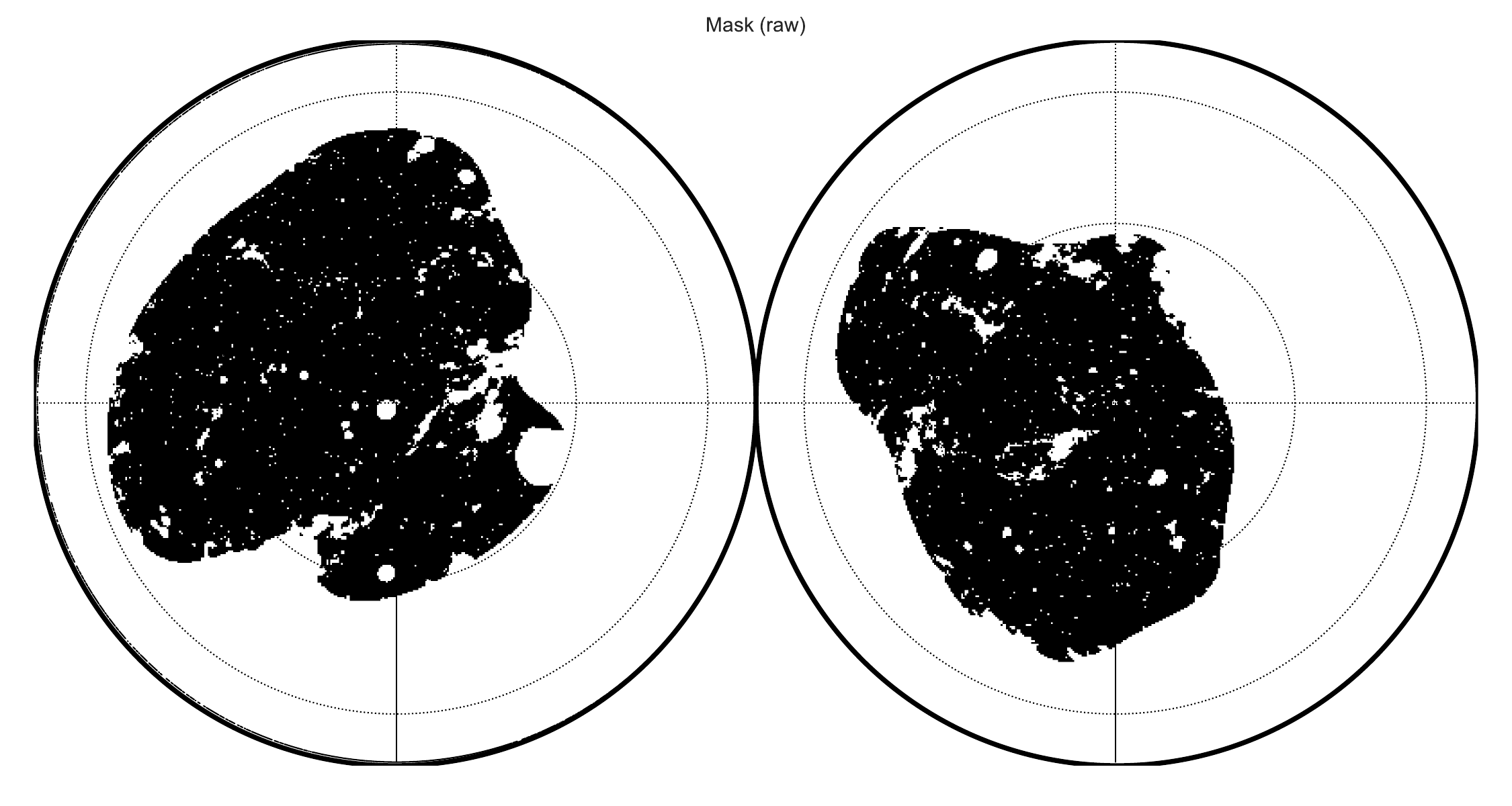}
    \includegraphics[width=\columnwidth]{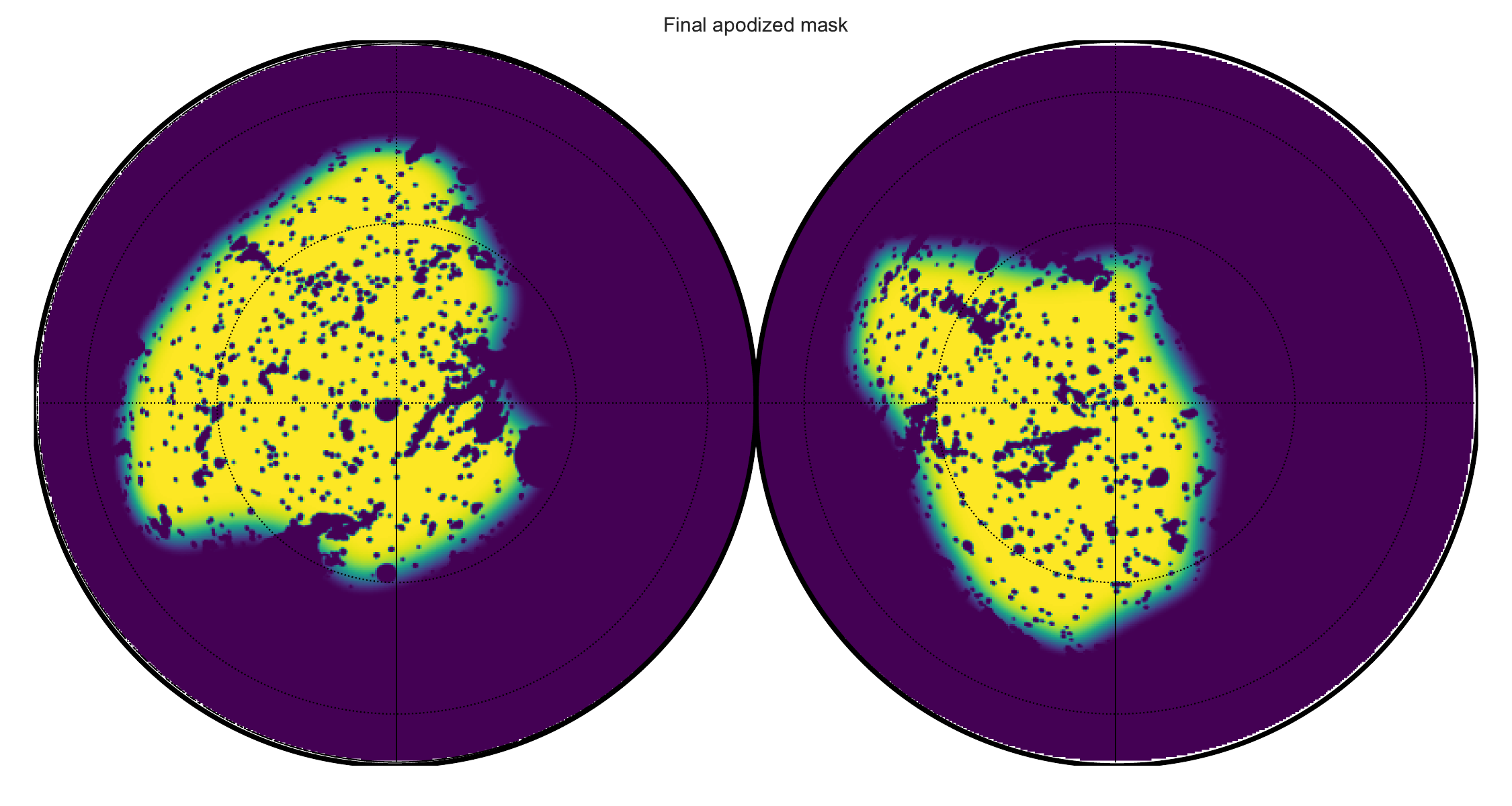}
	\caption{Sky mask used for the CIB analysis in orthographic projection, with the Galactic poles at the center (north pole left, south pole right). This figure shows the mask for the 545\;GHz data, but the differences from the other frequency masks are very minor and result only from the masking of non-CIB residuals and the point sources. \textbf{Top:} the boolean mask, showing the absence of the Galactic plane, the filamentary structure where high \ion{H}{i} column densities ($N_{\ion{H}{i}} > 2.5\times 10^{20}\,\rm cm^{-2}$) are masked, smaller point sources, and large regions that correspond to molecular intermediate-velocity clouds. \textbf{Bottom:} the same mask after apodization.}
	\label{fig:masks}	
\end{figure}

To compute the power spectra of our maps more reliably, we apodize these masks following the procedure for cosine apodization described in the \texttt{NaMaster} software \citep{alonso2018}. We choose an apodization kernel with an FWHM of $15'$ for this procedure. The final boolean mask and the apodized mask are presented in Figure \ref{fig:masks}.

\begin{figure*}
	\includegraphics[width=\columnwidth]{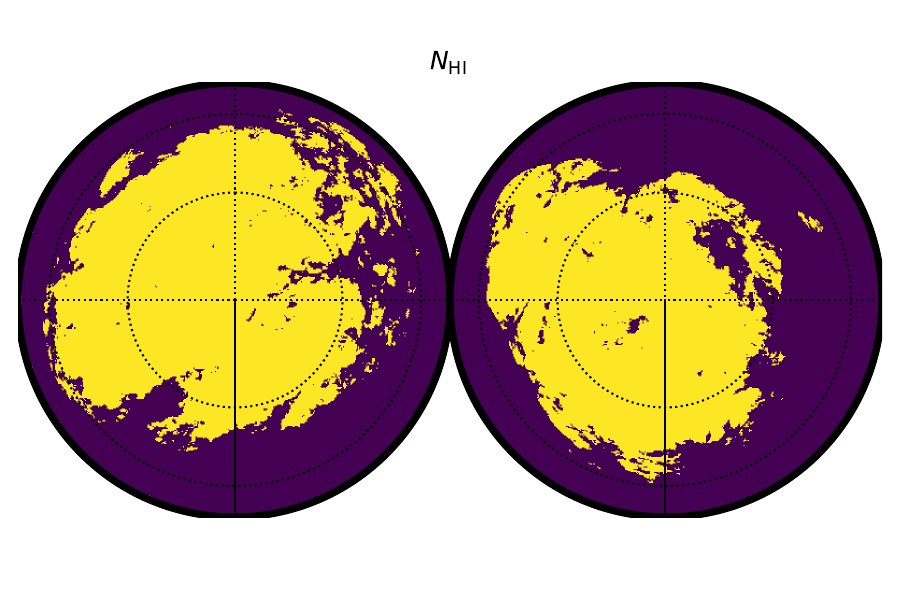}
    \includegraphics[width=\columnwidth]{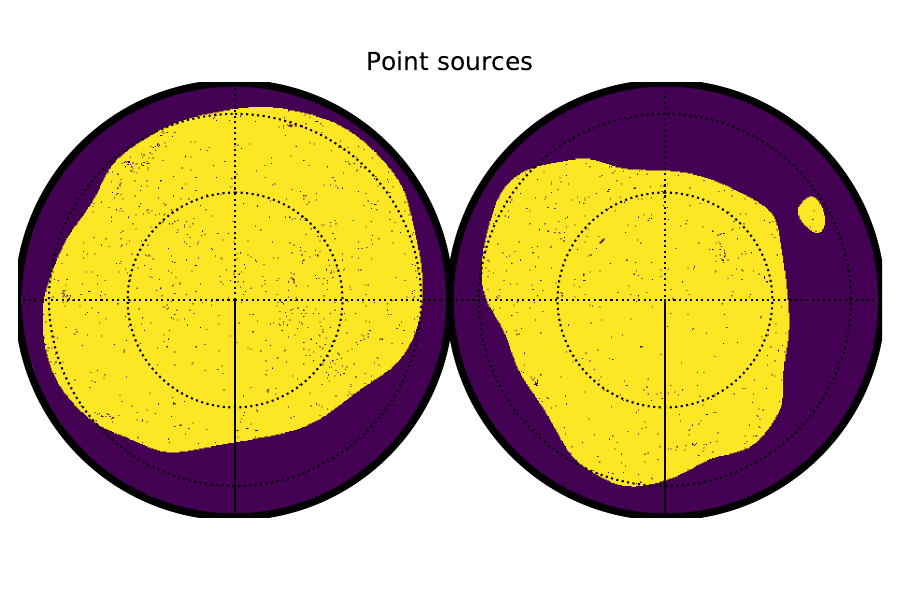}
	\includegraphics[width=\columnwidth]{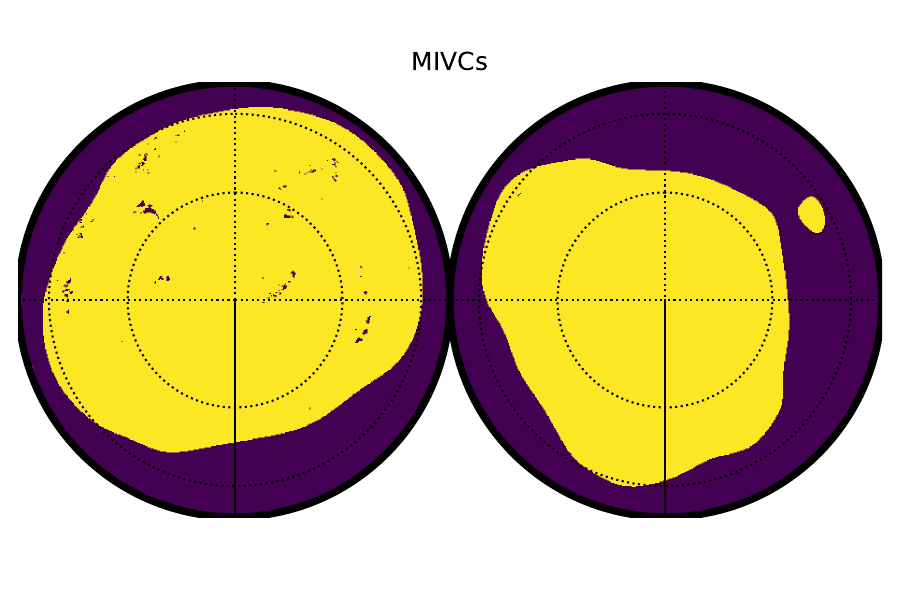}
    \includegraphics[width=\columnwidth]{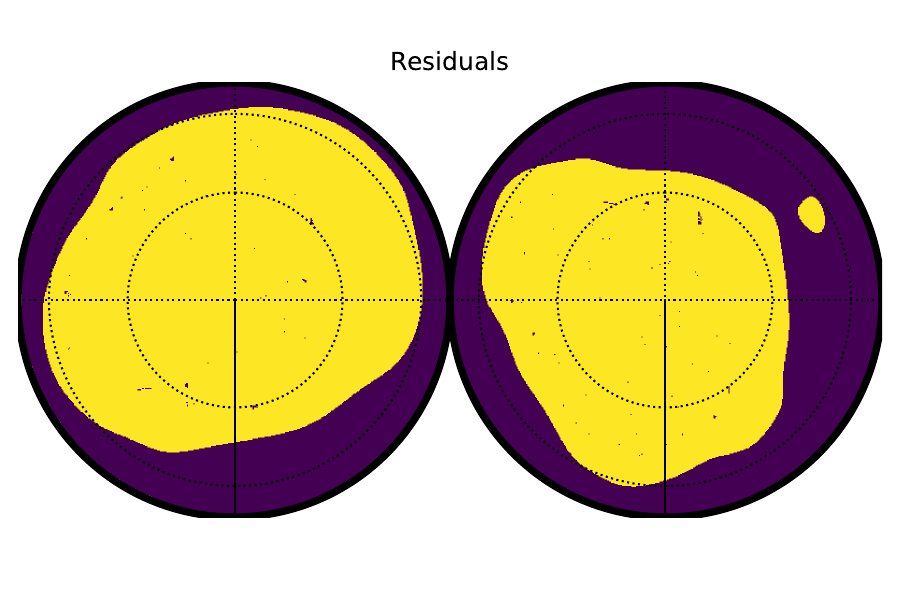}
	\caption{The different components that contribute to the total mask. $N_{\ion{H}{i}}$ is the \ion{H}{i} column density limit (here for $N_{\ion{H}{i}} < 3.0\times 10^{20}\,\rm cm^{-2}$ for illustrative purposes), the point sources are taken from the public \planck point source mask (given for each frequency), and the mask for the MIVCs is taken from \citet{roehser2016}. The mask of residual Galactic dust is a product of our iterative outlier masking scheme, further described in Section \ref{sect:glm_method}. All these masks are furthermore constrained by the \planck 20\% Galactic plane mask.}
	\label{fig:mask_components}	
\end{figure*}


\section{Building new CIB maps}
\label{sect:methods}

To disentangle the FIR emission from dust in the Milky Way and that from unresolved CIB galaxies, two different methods have been effectively used to date. We describe them in the following sections. For completeness, we note that the spectral information from the FIR signal alone cannot be utilized to separate Galactic and CIB dust, unlike what is done to separate the CMB component \citep[e.g. Commander, ][]{eriksen2008, planck2016_x}. This is due to the fact that both dust components are very well described by a modified blackbody, and a component separation based on this information alone would thus be strongly degenerate.

\subsection{Disentangling via the spatial structure}
\label{sect:spatial_structure}

The different spatial structures of the CIB and Galactic dust can be used to disentangle the two. This has been demonstrated with an implementation of the GNILC method \citep{planck2016_xlviii} or by subtracting the foreground dust at the power spectrum level \citep{mak2017}.

Using the differences in angular power spectra of these two components introduces a powerful source of information, but is also limited when the power spectrum of the fields that are reconstructed is not known to some degree. \citet{mak2017} assume a parametric form for the CIB power spectrum, which has been previously derived through \ion{H}{i}-based component separation \citep[][hereafter P14]{planck2014_xxx}. An approach based on the GNILC method \citep{planck2016_xlviii} has shown that Galactic dust and CIB emission can be disentangled for large parts of the sky ($\gtrsim 60\%$), albeit with the caveat of oversubtracting the CIB \citep[see][Appendix A]{maniyar2019}.

\subsection{Disentangling via the \ion{H}{i} column density}
\label{sect:nhi_method}

A commonly used approach to remove the foreground FIR intensity is to model its value based on the \ion{H}{i} column density \citep[e.g.][]{planck2011_xviii, planck2014_xxx}. In this case we model the observed FIR intensity in the following way:
\begin{equation}
    I_{\nu}(\alpha,\delta) = \sum_i \alpha^i_{\nu}\cdot N^i_{\ion{H}{i}}(\alpha,\delta) + \beta_{\nu}
    \label{eq:standard}
\end{equation}
Here, $I_{\nu}$ is the observed FIR intensity at frequency $\nu$, $\alpha^i_{\nu}$ is the dust emissivity per hydrogen nucleon, $N^i_{\ion{H}{i}}$ is the \ion{H}{i} column density, and $\beta_{\nu}$ is the zero-point. The subscript $\nu$ indicates that this equation is solved at each FIR frequency individually and the arguments $(\alpha, \delta)$ indicate the position on the sky. The index $i$ indicates that the spectroscopic \ion{H}{i} data can be binned into several column density maps with different velocity ranges. This approach is motivated by the fact that the dust emissivity $\alpha^i_{\nu}$ is different for multiple components because of a different composition or illumination by the interstellar radiation field (ISRF). By accounting for this difference, the FIR intensity can be modeled more accurately. The offset $\beta_{\nu}$ is required to better capture the spatially varying zero-point of dust and gas. In particular, the warm ionized medium \citep[e.g.][]{lagache1999} contributes to large-scale fluctuations of this zero-point.

Most commonly, we distinguish between low-velocity, intermediate-velocity, and high-velocity clouds \citep[LVCs, IVCs, and HVCs; see the review by][and references therein]{putman2012}. Usually, the shape of the \ion{H}{i} spectrum is used to define these velocity ranges. An illustration of this procedure can be found in Figure 1 of \citet{planck2011_xxiv}. In practice, the separation of the \ion{H}{i} data cube into multiple column density maps is done manually and can be subjective. Due to these limits, it is difficult to apply this method over areas of thousands of square degrees.

The most recent applications of this approach, particularly in light of the \planck data, are demonstrated in \citet{planck2011_xviii} and \citet{planck2014_xxx}. Here, \ion{H}{i} data from the GHIGLS survey \citep{martin2015}, from GASS \citep{kalberla2015}, and from EBHIS \citep{winkel2010} have been used to clean the FIR intensity maps of foreground dust.

Similar to many previous studies on the subject \citep[e.g.][for the latest application of this approach on FIR data]{planck2014_xxx}, we use the \ion{H}{i} data as a template for Galactic dust. However, to extend the sky coverage we use a slightly more automated and general approach. In the following, we describe the details of our component separation technique.

We focus on two aspects. First, we describe the spatial selection of the regions where the Galactic dust and CIB emission will be separated. The challenge here is to account for the spatial variability of the dust-to-gas ratio, while at the same time preserving the large-scale CIB fluctuations. Second, we detail how we utilize the full three-dimensional \ion{H}{i} data set, and how the information on the radial velocity of the gas helps to optimally construct a model of Galactic dust.

\subsection{Selecting the spatial structures}

The correlation between gas and dust varies across different physical environments and thus across the sky. Thus, we divide the sky into patches and perform the foreground modeling on one patch at a time, allowing us to turn this into an embarrassingly parallel problem. This procedure has already been proven to work very effectively in \citet{planck2014_xxx} and in \citet{planck2014_xviii}.

\begin{figure}
	\includegraphics[width=\columnwidth]{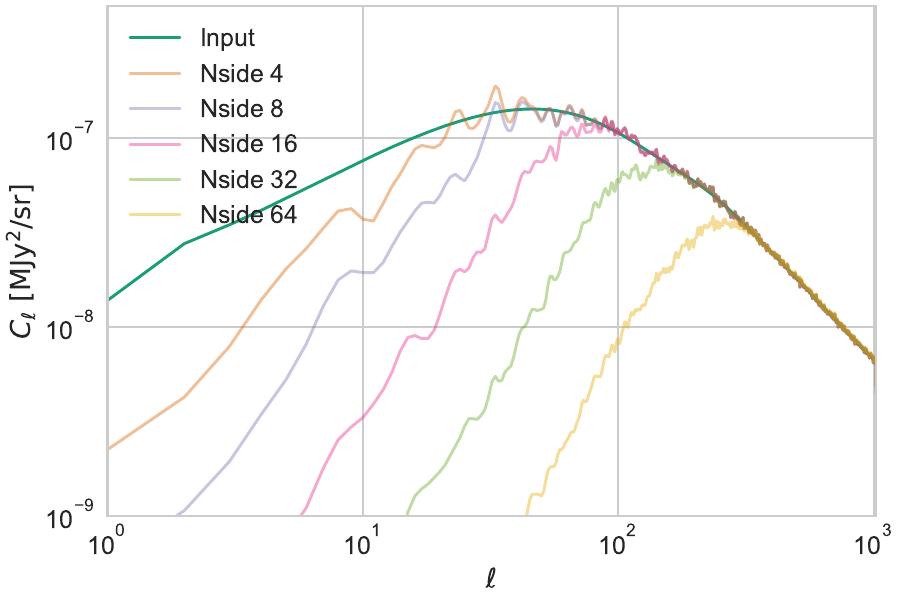}
	\caption{Effective high-pass filtering of the CIB auto power spectrum by working on individual patches. The different reconstructions use different HEALPix Nsides to split the sky into patches. For each patch, the mean is subtracted. This does not alter the CIB map visually or on small scales in the power spectrum, but removes most of the large-scale power.}
	\label{fig:hipass_filtering}	
\end{figure}

One caveat to this procedure is that a \textit{local} determination of the dust-to-gas ratios, and through that the CIB intensity, removes large-scale fluctuations of the CIB that extend beyond the size of the patches. In the power spectrum domain, we observe this as a sharp drop at low $\ell$. This effect has also been observed in \citet[][Figure 7, right panel]{planck2014_xviii} and \citet{schmidt2015}. We show an example of this effect in Figure \ref{fig:hipass_filtering}. There, we generate a simulated CIB map at 545\,GHz and subtract the mean value in each patch. For the patches, we use the HEALPix pixelization and demonstrate the effect for different Nsides. In the following, we refer to these larger HEALPix patches as superpixels.

Due to this limitation, we set up our procedure as follows. The HEALPix pixels on which we locally perform the modeling are chosen to have an Nside of 16 (we further discuss this choice in Section \ref{sect:validation:superpix}), giving us 4096 full-resolution pixels in one patch. After applying the model that is described in more detail in the following section, we convolve the full-sky maps of model parameters with a Gaussian kernel with an FWHM of $3^{\circ}$, thereby creating continuous maps of dust-to-gas ratios for different radial velocities, and for the zero-point in the \ion{H}{i}/FIR relation. A similar version of this procedure has already been applied by \citet{planck2014_xxx} in their analysis of the large GASS field.

To generate the dust model, these continuous parameter maps are then combined with the \ion{H}{i} data (Eq. \ref{eq:glm}), which are subject to this spatial filtering. We further discuss the implications of this in Section \ref{sect:validation:superpix}.

\subsection{Using the full spectral \ion{H}{i} information}
\label{sect:glm_method}

A common practice is to manually set the velocity ranges of the different \ion{H}{i} column density maps for a single field. There are three caveats to this procedure. First, the separation is manual and subject to human opinion and error. Second, the typical number of column density maps for one field ranges from one to three, thus offering no possibility to account for higher-resolution substructures. Third, the separation may not be well defined for a given field, allowing different interpretations of where the low/intermediate/high-velocity gas intersections are.

To overcome this manual step and to allow the model to be more universal, we implement a generalized linear model (GLM), similar to the analysis done in \citet{lenz2016}. This means that we assume different emissivities for different radial velocity bins by allowing each spectral channel to have an individual dust emissivity. Building up on Eq. (\ref{eq:standard}), the GLM for the FIR intensity $I_{\nu}(\alpha,\delta)$ at sky coordinates $(\alpha,\delta)$ can be written as a weighted sum over all spectral channels:
\begin{equation}
    I_{\nu}(\alpha,\delta) = \sum_{\mathrm{ch}} \epsilon^{\mathrm{ch}}_{\nu}\cdot T_B^{\mathrm{ch}}(\alpha,\delta) + \beta_{\nu}.
    \label{eq:glm}
\end{equation}
%

Here, $T_B^{\mathrm{ch}}(\alpha,\delta)$ denotes the \ion{H}{i} brightness temperature in each spectral channel and $\epsilon^{\mathrm{ch}}_{\nu}$ is the emissivity for each individual channel. Similar to Eq. (\ref{eq:standard}), $\beta_{\nu}$ is a constant offset.

The HI4PI data have a total of 933 spectral channels with a channel width of $1.3\,\rm km\,s^{-1}$ each, which gives us far more spectral resolution than required. By spectrally binning the \ion{H}{i} as described in Section \ref{sect:data:hi4pi}, we still have a sufficient number of spectral \ion{H}{i} components while at the same time reducing the degeneracy of the model, reducing the computational cost, and increasing the sensitivity of the data.

Many of the \ion{H}{i} spectral channels contain no significant emission and are masked when solving Eq. (\ref{eq:glm}). We use the 3$\sigma_{\mathrm{RMS}}$ of the \ion{H}{i} data as a clip level. This avoids that CIB fluctuations being falsely fit by using a large number of degrees of freedom, thereby creating an artificial, false image of the background component.

To further constrain our model and avoid chance correlation of \ion{H}{i} and the CIB, we use lasso regularization for the individual GLM coefficients $\alpha^{\mathrm{ch}}_{\nu}$ \citep{tibshirani1996}. Thereby, we break the degeneracy between neighboring, correlated channels and avoid overfitting. In practice, this means that we define our cost function $J(\theta)$ as the sum of a square loss function $L(f(X|\theta), Y)$ and the total sum of the coefficients $C(\theta)$:
\begin{eqnarray}
	J(\theta) &=& L(f(X|\theta), Y) + C(\theta)\\
	L(f(X|\theta), Y) &=& ||I_{\nu} - \sum_{\mathrm{ch}} T_B^{\mathrm{ch}}\epsilon_{\nu}^{\mathrm{ch}}-\beta_{\nu}||^2\\
	C(\theta) &=& \gamma \sum_{\mathrm{ch}} || \epsilon_{\nu}^{\mathrm{ch}}||_1
    \label{eq:lasso}
\end{eqnarray}
Here, $||\cdot||_1$ denotes the $L^1$-norm and $\gamma$ is the strength of the regularization. The second term ensures that the fit yields a sparse set of coefficients.

To scale the regularization strength $\gamma$, we use cross validation \citep[e.g.][]{picard1984}. This technique is commonly used to optimize hyperparameters such as $\gamma$, which are not directly evaluated on the data, but need to be determined before the actual fitting procedure. For this purpose, the image is split into $n$ parts (so-called folds). For the $n$-fold cross validation, the data are fitted on $n-1$ folds (training sample) and the quality of the fit is evaluated on the $n$th fold (test sample). Each of these folds is randomly selected, and they do not represent spatially coherent features. This procedure is repeated $n$ times, so each fold serves as test sample exactly once. Therefore, the GLM is not evaluated just once for an image, but instead hundreds of times to find an appropriate solution. For our purpose, we work with values of $n = 3 ... 5$ and find that the exact choice of $n$ does not affect the results significantly.

\begin{figure}[tp]
	\includegraphics[width=\columnwidth]{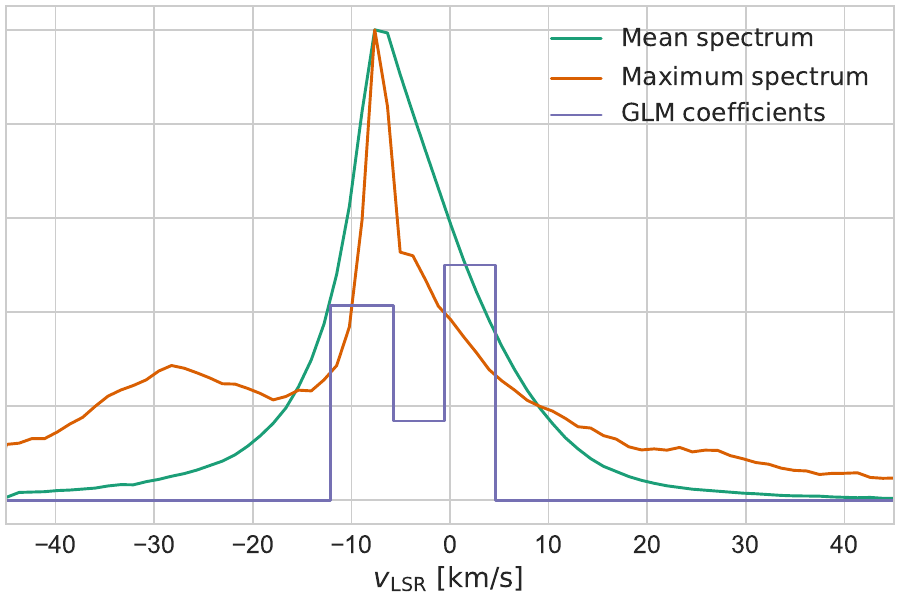}
	\caption{GLM coefficients $\epsilon_{\nu}^{\mathrm{ch}}$ (purple), mean \ion{H}{i} spectrum (green), and maximum \ion{H}{i} spectrum (orange) for a random superpixel. The GLM coefficients are normalized to the maximum spectrum and multiplied by -1 for illustration purposes. For the resulting map, see Figure \ref{fig:fitresults_maps}.}
\label{fig:spectrum}
\end{figure}

\begin{figure}[tp]
	\includegraphics[width=\columnwidth]{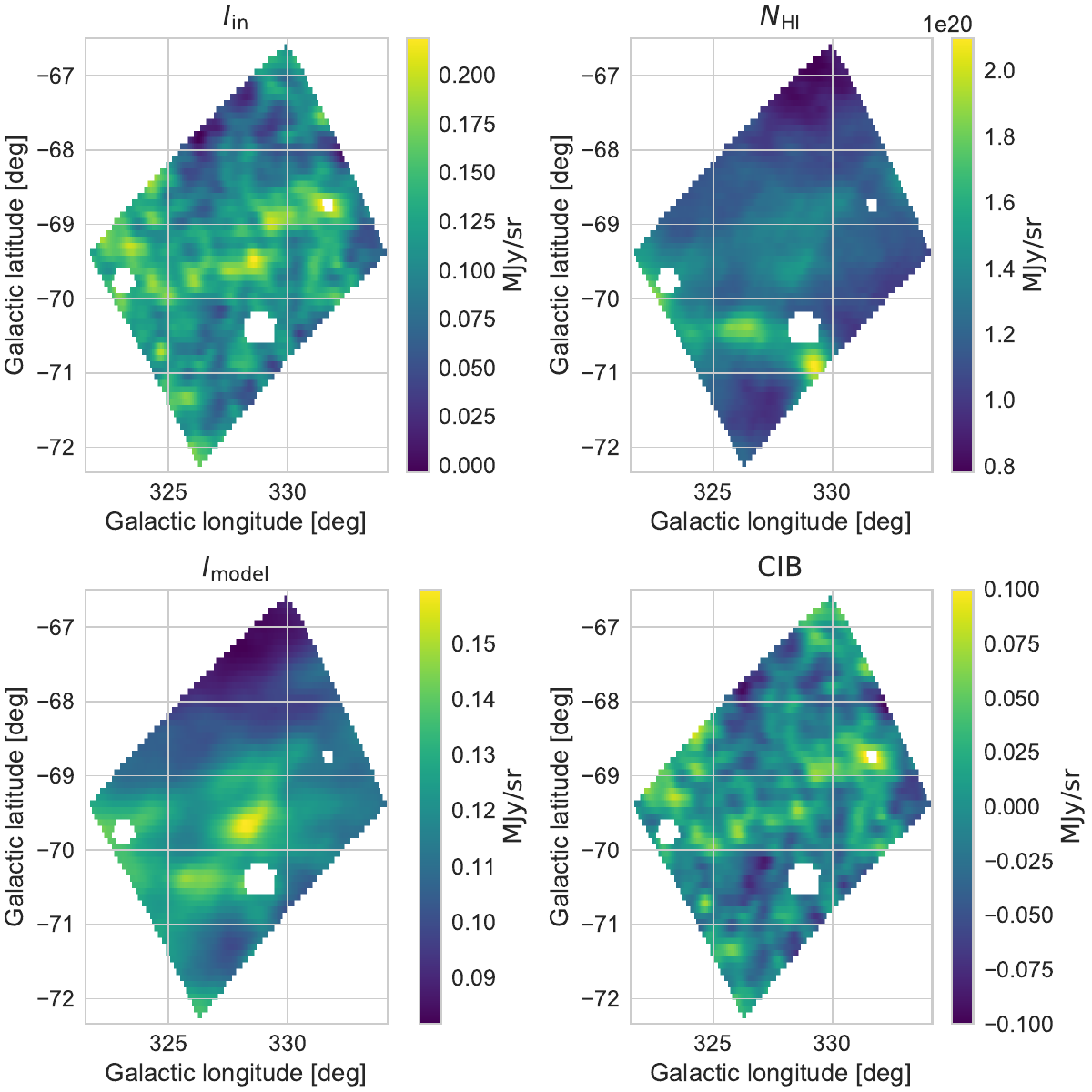}
	\caption{Illustration of the component separation matching Figure \ref{fig:spectrum}. \textbf{Top left:} input \planck intensity at 545\,GHz. \textbf{Top right:} total \ion{H}{i} column density. \textbf{Bottom left:} foreground dust model, based on the \ion{H}{i} data and the GLM. \textbf{Bottom right:} resulting CIB image.}
\label{fig:fitresults_maps}
\end{figure}

An illustration of the GLM approach is given in Figure \ref{fig:spectrum} for a random superpixel. We show the normalized GLM coefficients, the mean spectrum for that superpixel, and the maximum spectrum. The latter helps to identify spatially small features in an otherwise low-signal region. We further demonstrate the result of this technique in Figure \ref{fig:fitresults_maps}, where we show the FIR data, model, and derived CIB map.

To remove the residual emission resulting from CO-dark molecular gas, we repeat iteratively the procedure detailed above and evaluate the Gaussian distribution of the resulting CIB maps. Outliers of more than $3\sigma$ are masked and the modeling step is repeated until convergence is reached (see also Figure 12 of \citet{planck2011_xxiv} for further reference).

Finally, we obtain the CIB maps by subtracting the dust models from the total intensity maps.

We note that the discrete nature of the FIR emissivities that is assumed in this approach is primarily a schematic description and does not necessarily yield an accurate value of the physical dust emissivity for each binned spectral channel of the \ion{H}{i} data. This is acceptable, however, because our main goal here is to create a robust foreground model, free of residuals from Galactic dust.

\section{$C_{\ell}$ estimation}
\label{sect:powerspectra}

To estimate the angular power spectra, based on the CIB and CMB lensing fields, we use \texttt{PyMaster}, the Python implementation of \texttt{NaMaster}\footnote{\url{https://github.com/LSSTDESC/NaMaster}} \citep{alonso2018}. When estimating the $C_{\ell}$ at large scales, incomplete sky coverage leads to mode--mode coupling, often approximately described in the matrix $M_{\ell \ell'}$ \citep{hivon2002}. This effect is particularly strong at larger scales, and we found significant differences between the $C_{\ell}$ estimate using \texttt{healpy.anafast} (no correction for the mask) and that using \texttt{NaMaster} (partially correct for the mask), as expected.

We constrain the power spectra in the range $\ell < 2000$ for two reasons. First, the resolution of the CIB maps is 5', and any information above $\ell = 2000$ would be sensitive to inaccuracies in the beam correction. Second, the underlying \ion{H}{i} data that are used to build the dust model have a resolution of $16.2'$. Hence, dust residuals on very small scales cannot be removed by the low-resolution \ion{H}{i} data and contaminate the power spectrum. 

To avoid ringing at the smaller scales in the power spectrum, we apodize the mask as described in Section \ref{sect:data:masks}. To measure the CIB angular power spectra at a given frequency, we measure the cross power spectra between two CIB maps built on the two ring halves, eliminating the correlated noise term that would impact the smaller scales. The power spectra are furthermore binned with a bin factor of \nlb $\ell$ modes per bin. Lastly, we use the covariance of the power spectra as a measure of the uncertainty in our power spectra analysis. We present a more detailed discussion of this in Section \ref{sect:validation:cl_errors}. We validated the stage of estimation of the power spectrum extensively using simulations. 

\section{Map Results}
\label{sect:map_results}

In this section, we present the results exemplary for the 545\,GHz channel as an example. The corresponding results for the other frequencies can be found in Appendix~\ref{sect:app:maps}.

\begin{figure*}[tp]
	\includegraphics[bb=100 0 1000 600, clip=, width=\columnwidth]{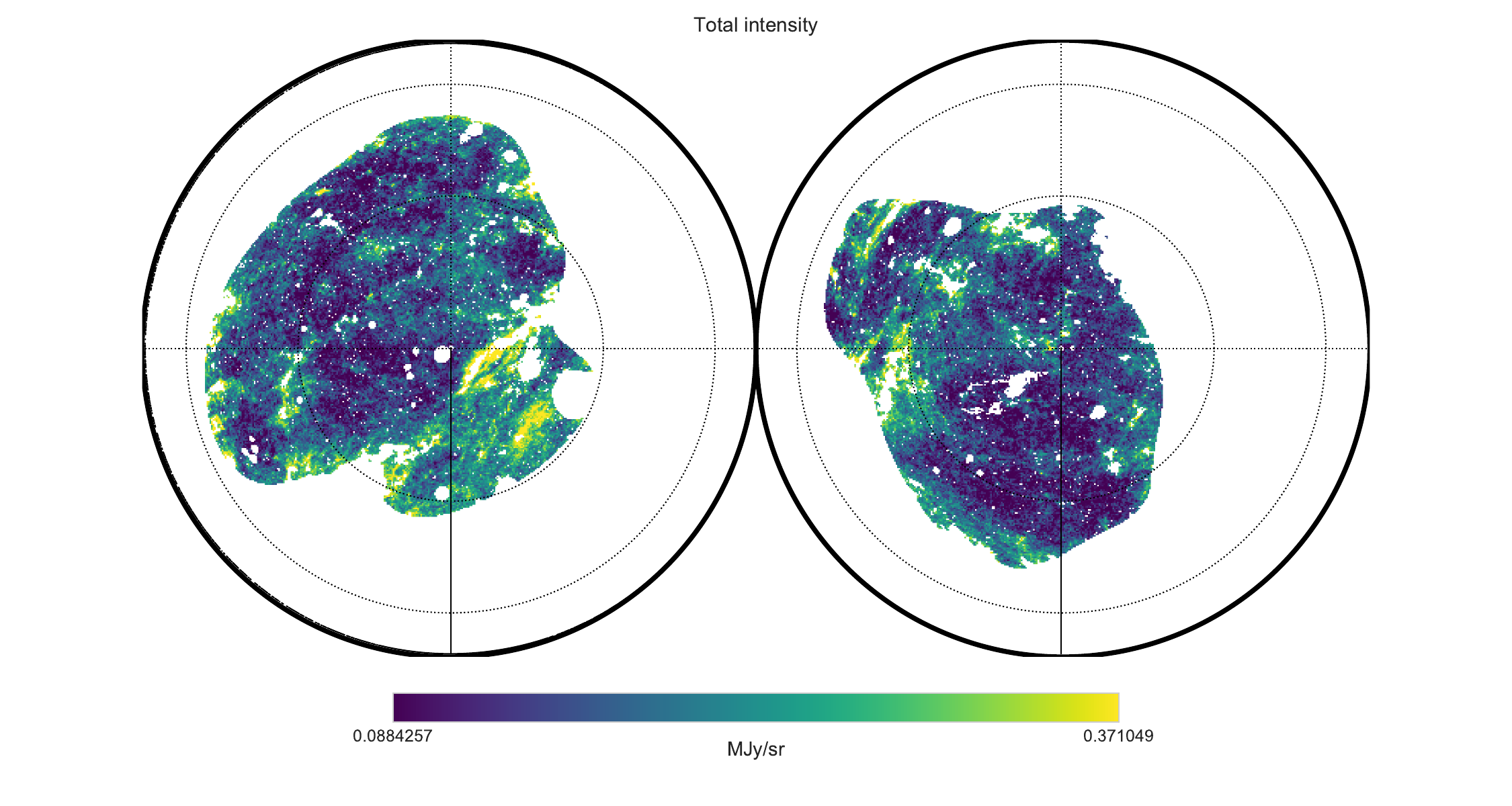}
	\includegraphics[bb=100 0 1000 600, clip=, width=\columnwidth]{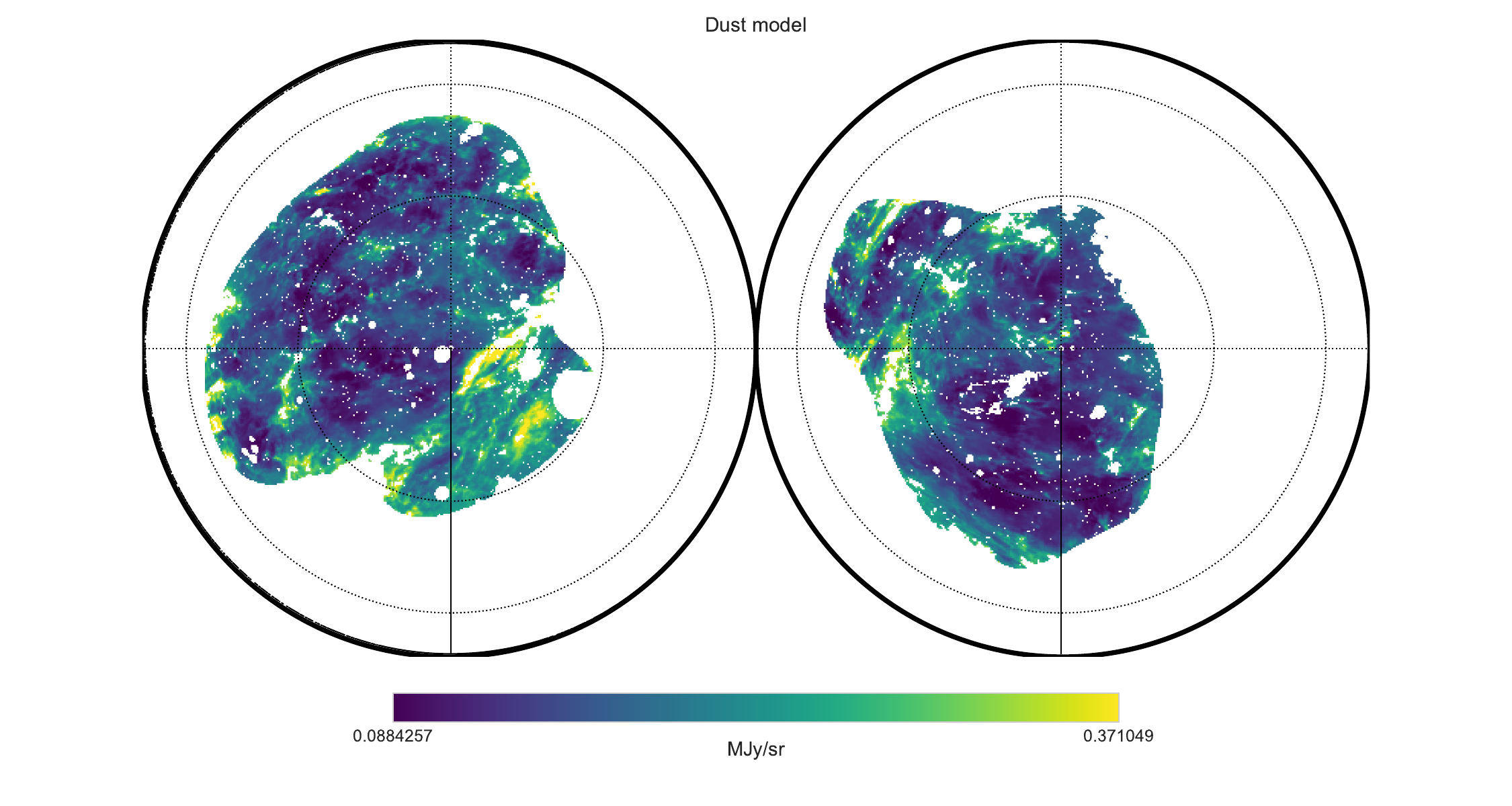}
	\caption{\textbf{Left:} total FIR intensity at 545\,GHz for the Galactic north pole (left) and south pole (right). At this frequency, the sky is dominated by Galactic dust and CIB emission. \textbf{Right:} \ion{H}{i}-based dust model for the Galactic north pole (left) and south pole (right). It can be seen that the small-scale CIB fluctuations are only present in the total FIR intensity map, and not in the dust model. Moreover, no sign of the underlying patches in which the foreground model was computed can be seen.}
\label{fig:orth_intensities}
\end{figure*}

We present the input FIR emission and the modeled foreground dust in Figure \ref{fig:orth_intensities}, which nicely shows how the \ion{H}{i}-based dust model captures the large-scale cirrus features, while not picking up the CIB fluctuations that can be seen in the intensity map. Moreover, no imprint of the underlying patches in which the dust model is computed is visible.

\begin{figure*}[tp]
	\includegraphics[bb=0 0 600 400, clip=, width=\textwidth]{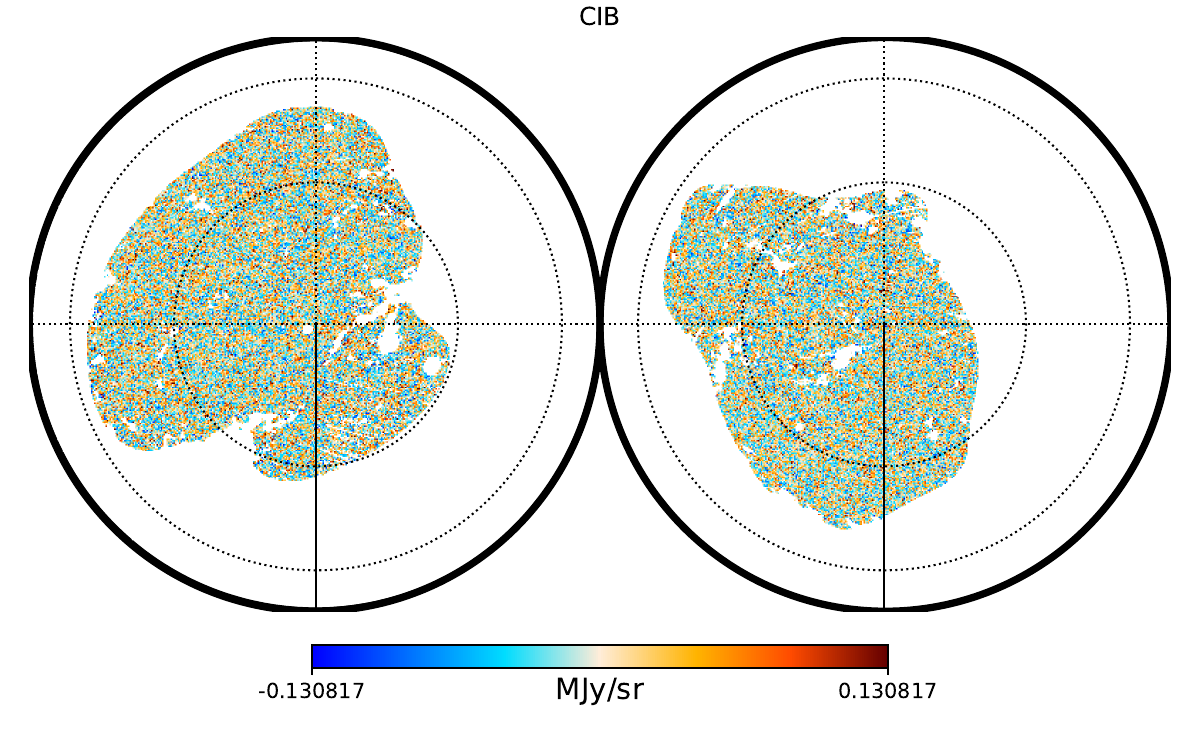}
	\caption{CIB anisotropies at 545\,GHz for the Galactic poles (north left, south right). This footprint covers \fsky\% of the sky.}
\label{fig:orth_cib}
\end{figure*}

Figure \ref{fig:orth_cib} presents the large-scale, full-resolution 545\,GHz map. This is based on the full mission data and has a resolution of $5'$. This map covers a total of $\fsky\%$ of the sky.

\begin{figure}[tp]
	\includegraphics[width=\columnwidth]{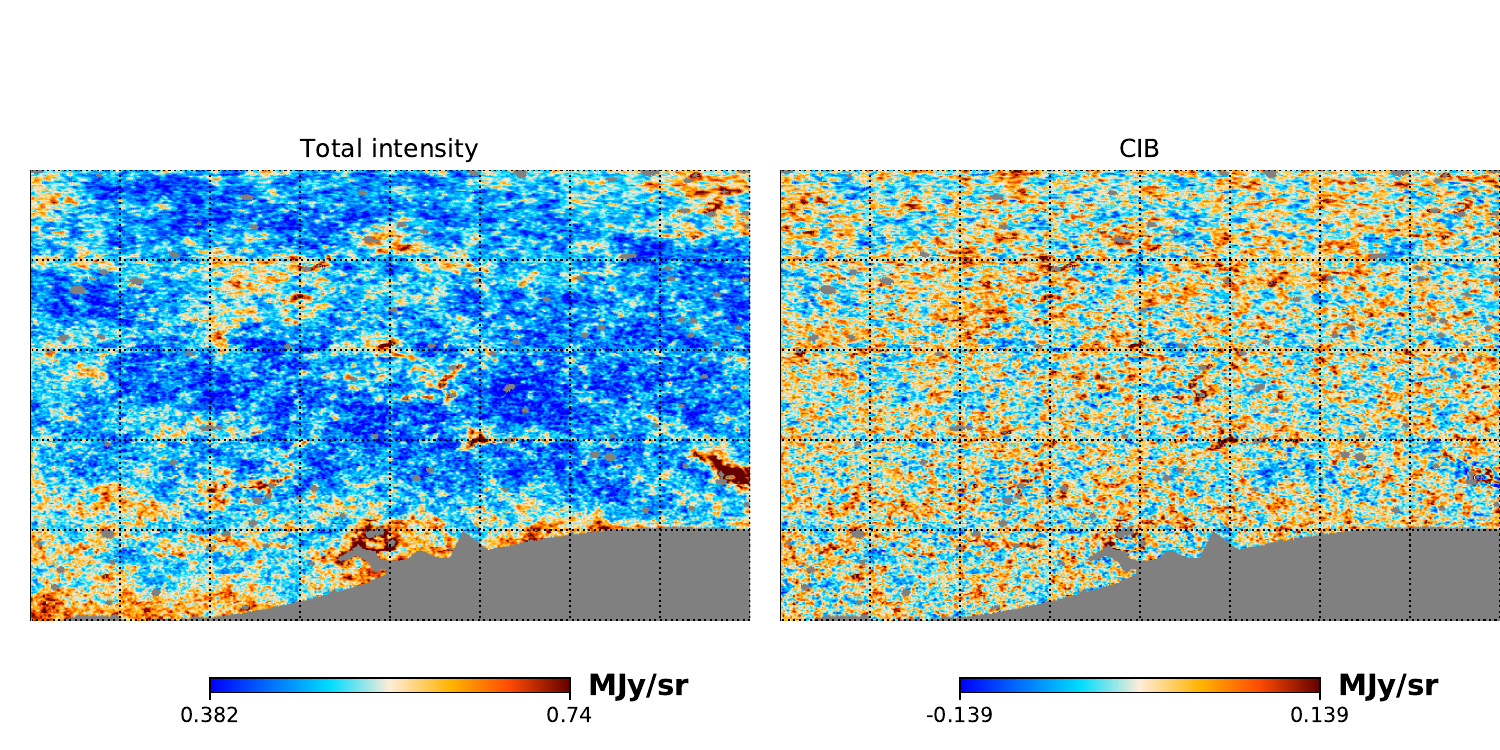}
	\caption{Comparison of total FIR intensity (\textbf{left}) and the resulting CIB fluctuations (\textbf{right}) for a $40^{\circ}\times 25^{\circ}$ field. Almost all of the Galactic dust emission is removed, while no imprint of the underlying patches in which this subtraction is performed can be seen.}
\label{fig:smallscale_comp}
\end{figure}

A smaller sky region that shows the total FIR intensity and the CIB signal after foreground subtraction is presented in Figure \ref{fig:smallscale_comp}.

\begin{figure}[tp]
	\includegraphics[bb=100 0 1000 600, clip=, width=\columnwidth]{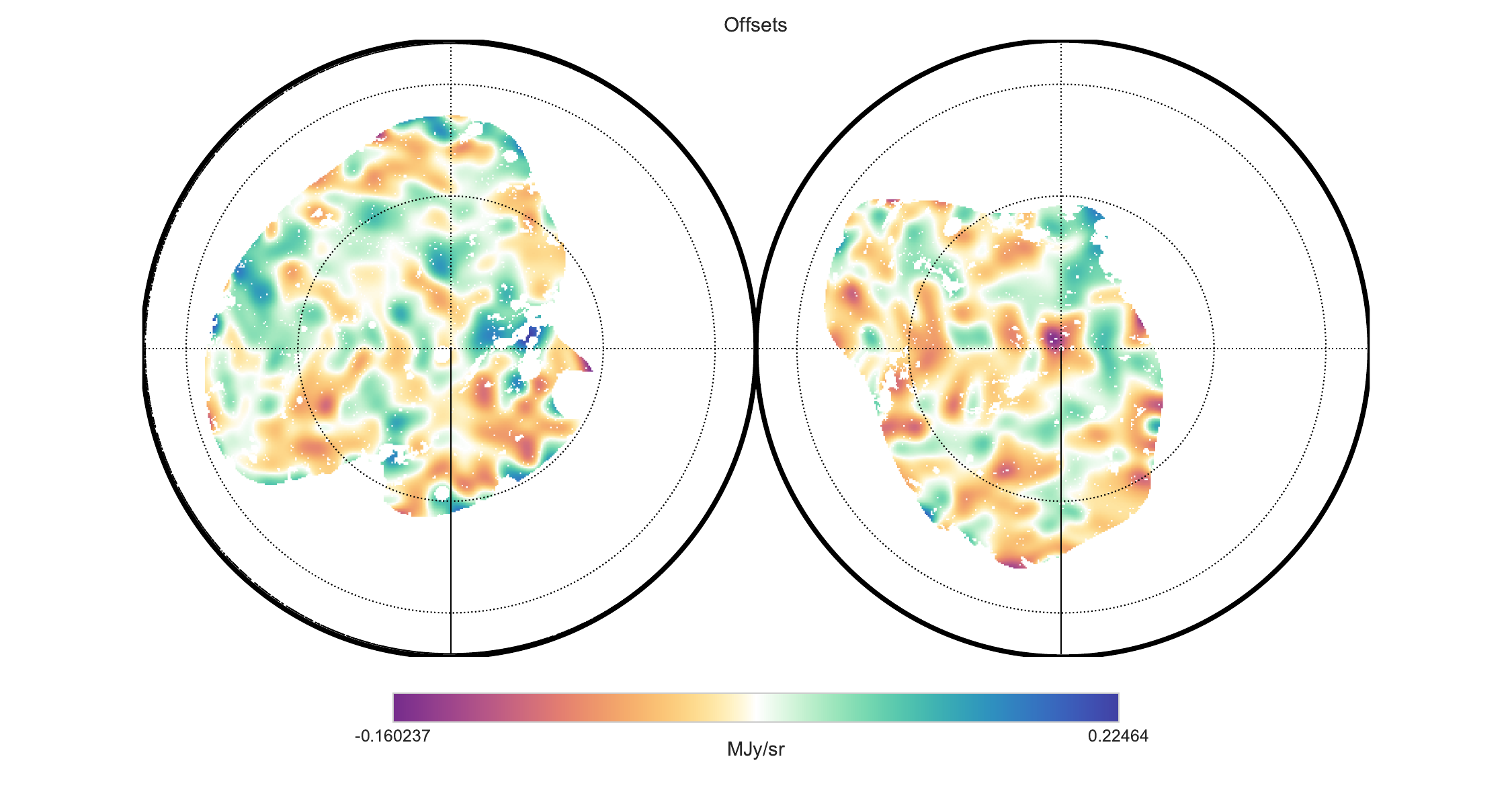}
	\caption{Smoothed offsets in the \ion{H}{i}--FIR relation, corresponding to $\beta_{\nu}$ in Eq. (\ref{eq:glm}). These large-scale variations of the offset are important to capture, otherwise the foreground modeling will fail (see Section \ref{sect:validation:offset}).}
\label{fig:offsets}
\end{figure}

As described in Section \ref{sect:methods}, we convolved the parameter maps with a $3^{\circ}$ Gaussian kernel to avoid any edge effects in the dust model. An example of such a smoothed parameter map is given in Figure \ref{fig:offsets}, where we show the spatially varying offset $\beta_{\nu}$ (see Eq. (\ref{eq:glm})). We note that the spatially varying offset is a critical parameter, without which the foreground model fails to provide us with an accurate CIB map (see also Section \ref{sect:validation:offset}).

\section{Results for the Power Spectra}
\label{sect:powerspectra_results}

In the following, we describe the results we have obtained from the angular power spectra of the CIB $C_{\ell}^{\nu_1\nu_2}$, and from the CIB--CMB lensing cross correlation $C_{\ell}^{T\kappa}$. The formalism to obtain these power spectra has been described in Section \ref{sect:powerspectra}.

\subsection{CIB auto power spectra}

\begin{figure*}[tp]
	\includegraphics[width=\textwidth]{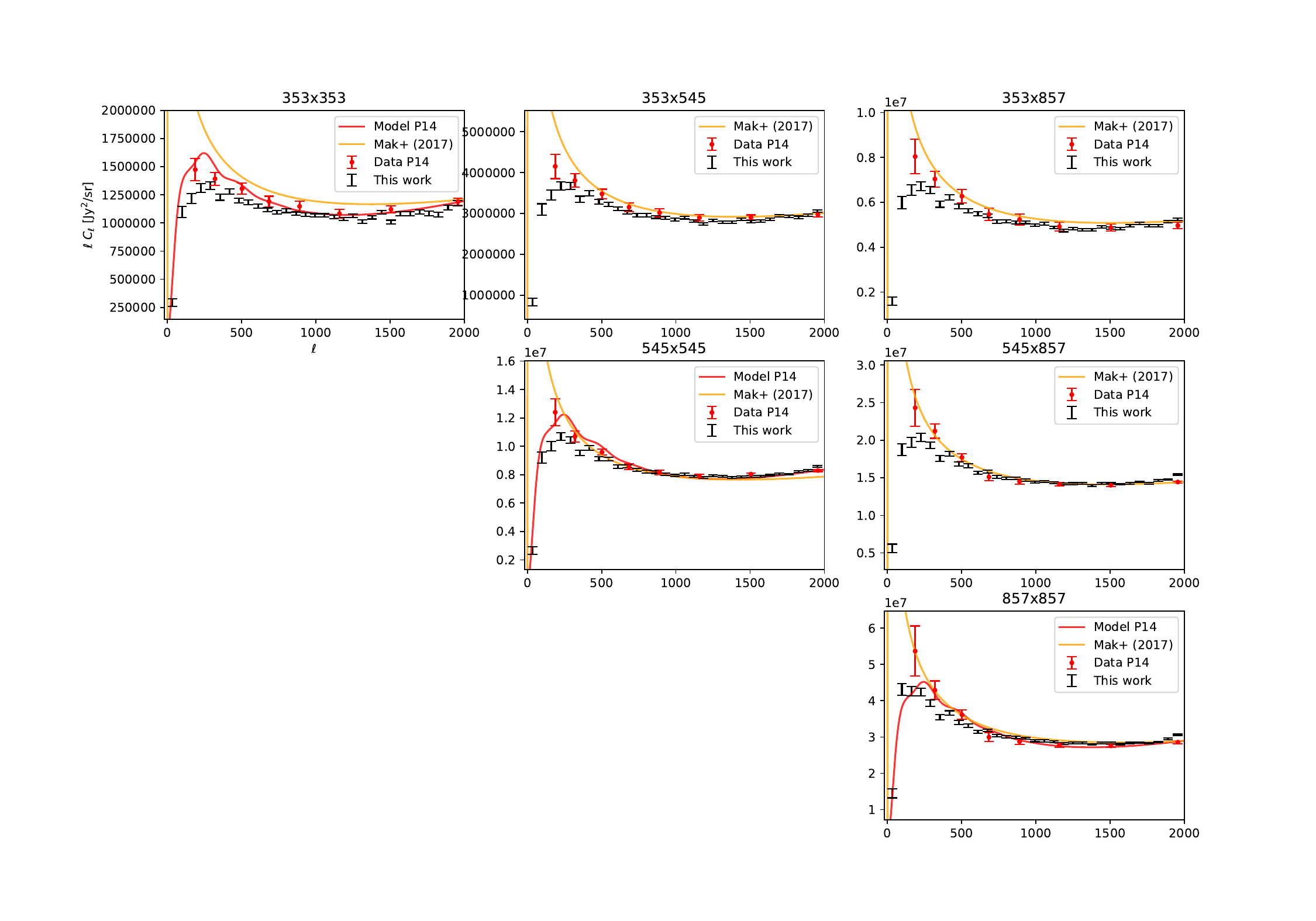}\\
	\caption{CIB auto and cross power spectra for the different frequencies (left to right: 353, 545, and 857\,GHz). For the auto power, we use the cross correlation of different ring halves for which the dust model was subtracted separately. The reference data and model (red error bars and red line) are taken from \citet{planck2014_xxx}, and the second reference model (yellow line) is taken from the power spectrum-based component separation presented in \citet{mak2017}. To see how the level of dust contamination varies with the selected sky fraction, see Figure \ref{fig:validation_nhi_thresholds_cibxcib}.}
\label{fig:cib_autopower}
\end{figure*}

The resulting CIB angular auto and cross power spectra $C_{\ell}^{\nu_1\nu_2}$ for the different frequencies are shown in Figure \ref{fig:cib_autopower}. We include a comparison to the data points and model of \citet{planck2014_xxx}, as well as the CIB extracted from the model presented in \citet{mak2017}. We did not correct the results that are based on previous \planck releases for the slightly different calibration, because these are not explicitly studied and are only of the order of $\sim 1-2\%$.

We find an excellent agreement with these previous studies in general, but several differences are worth discussing. First, we observe that \citet{mak2017} find more power on large scales than both \citet{planck2014_xxx} and the results presented here. This excess signal hints at residual contamination by Galactic dust, and could be the result of the power spectrum-based dust modeling in \citet{mak2017}, as opposed by the map-based dust removal here and in \citet{planck2014_xxx}.

Second, toward smaller scales and toward higher frequencies, we find a slightly higher CIB level than what is found in \citet{planck2014_xxx}. Despite the steep power law that describes the angular power spectrum of Galactic dust \citet[][$\propto \ell^{-2.7}$]{miville-deschenes2002}, we show in Section \ref{sect:validation:nhi_thresh} that this excess results from dust contamination.

For all CIB cross power spectra, we also estimate the cross-correlation coefficient spectrum $\rho_{\ell}^{\nu_1\,\nu_2}$, which is given by
\begin{equation}
	\rho_{\ell}^{\nu_1\,\nu_2} = \frac{C_{\ell}^{\nu_1\,\nu_2}}{\sqrt{C_{\ell}^{\nu_1\,\nu_1}\;C_{\ell}^{\nu_2\,\nu_2}}}
\end{equation}

\begin{figure}[tp]
	\includegraphics[width=\columnwidth]{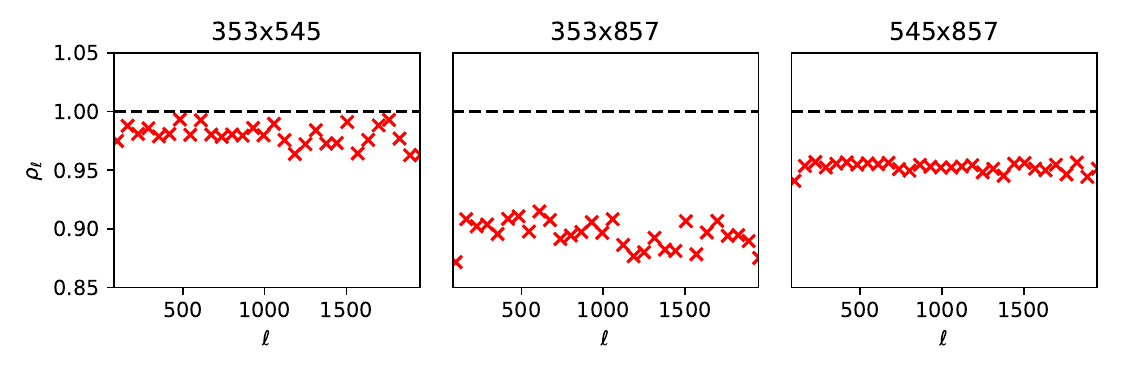}
	\caption{CIB cross correlation coefficients $\rho_{\ell}^{\nu_1\,\nu_2}$. We find that the correlation is scale-independent and, as expected, is strongest for neighboring frequency channels.}
\label{fig:cib_correlation_coeffs}
\end{figure}

The resulting plot is shown in Figure \ref{fig:cib_correlation_coeffs}. We furthermore compute the correlation coefficients averaged for the multipole range $150<\ell<1000$ (Table\,\ref{tab:cib_correlation_coeffs}). The values are in line with the ones presented in \citet[][their Table 11]{planck2014_xxx} and with \citet[][their Table 5]{mak2017}, which is expected given the agreement of the individual power spectra.

\begin{table}[h!]
\renewcommand{\thetable}{\arabic{table}}
\centering
\caption{CIB cross correlation coefficients, averaged over the multipole range $150<\ell<1000$}
\label{tab:cib_correlation_coeffs}
\begin{tabular}{ccccc}
\hline
& 353\,GHz & 545\,GHz & 857\,GHz\\
\hline
353\,GHz & 1 & $0.98\pm0.01$ & $0.91\pm0.01$ \\
545\,GHz & & 1 & $0.96\pm0.01$\\
857\,GHz & & & 1\\
\hline
\end{tabular}
\end{table}

\subsection{CIB--CMB lensing cross power spectra}

\begin{figure*}[tp]
	\includegraphics[width=\columnwidth]{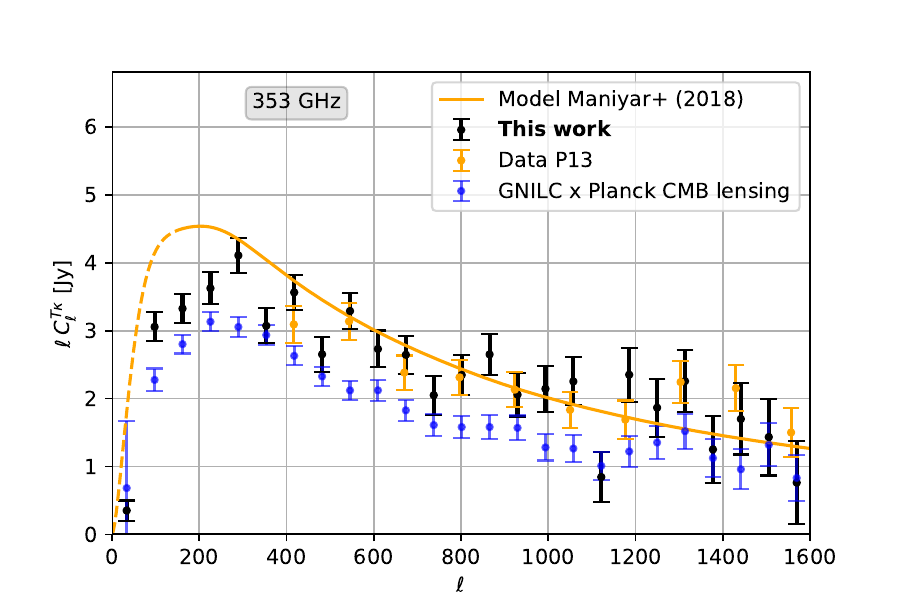}
	\includegraphics[width=\columnwidth]{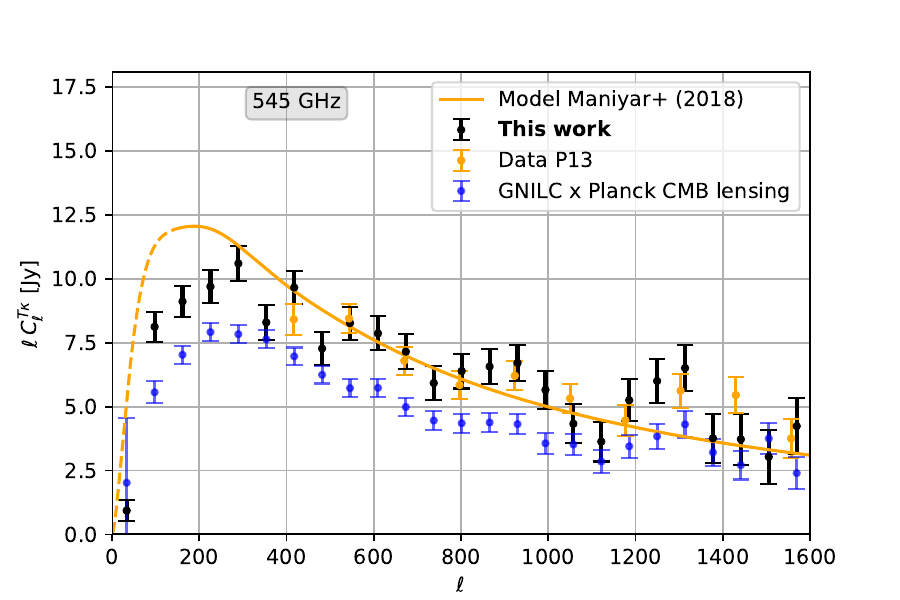}
	\includegraphics[width=\columnwidth]{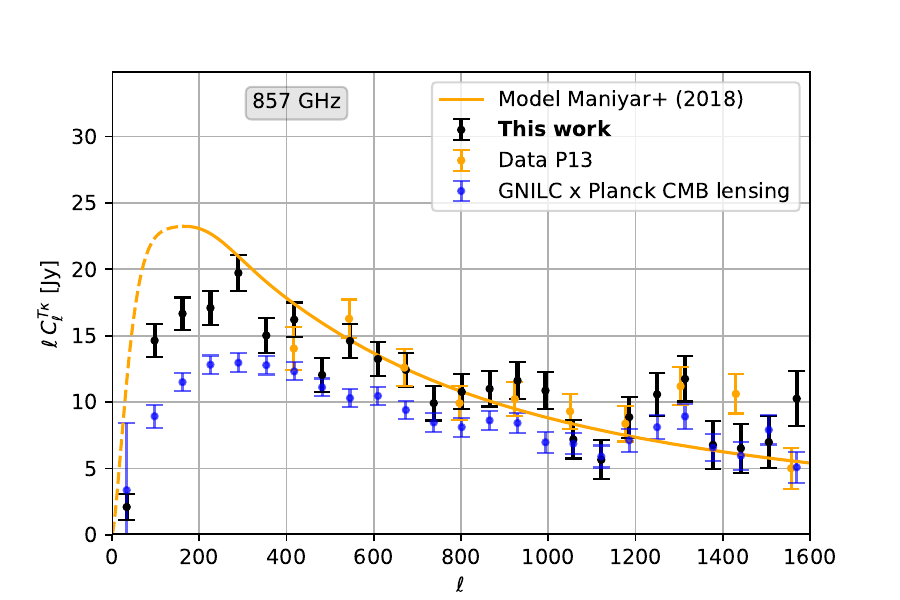}
	\caption{Cross power spectrum of the CIB and the lensing convergence $\kappa$ for the different \planck frequencies. Shown here are the results for an $N_{\ion{H}{i}}$ threshold of $2.5\times 10^{20}\;\rm cm^{-2}$. We find excellent agreement with the work presented in \citet{planck2014_xviii} and \citet{maniyar2018}, and extend the constraints on these cross power spectra to larger scales than previously probed. We also note that the CIB maps from \citet{planck2016_xvii} show a weaker cross correlation with the CMB lensing signal (see the text for more details).}
\label{fig:cl_cibxkappa}
\end{figure*}

The cross correlation of the CIB with the CMB lensing signal is a powerful tool to test the accuracy of the CIB maps, because any missing CIB flux will directly result in a weaker cross correlation. Furthermore, these two quantities are derived completely independently and their comparison serves as a good systematic and quality check. We present the result of this analysis in Figure \ref{fig:cl_cibxkappa}.

We find that our results on the CIB--CMB lensing cross correlation are in good agreement with the results obtained in \citet{planck2014_xviii}, and that the error bars are comparable as well for the higher frequencies. Consequently, our results are also in good agreement with the model presented in \citet{maniyar2018}, which is fit to the data from \citet{planck2014_xviii}. On top of that, this model uses the CIB auto power spectra and measurements of the CIB mean brightness from \citet{planck2014_xxx}. These data sets are then used to constrain the cosmic star formation history, the redshift-dependent bias, and the mass of galaxies that contribute most to the CIB emission. We also note that we omitted the two data points at the largest scales from \citet{planck2014_xviii} because they are only upper limits. Hence, the model of \citeauthor{maniyar2018} is also only mildly constrained on scales $\ell \lesssim 400$.

The dust cleaning allows us to access the large scales at which the cirrus contamination makes any insights difficult. Interestingly, we find that the peak of the power spectrum is at smaller scales than found previously \citep{planck2014_xviii, maniyar2018}, and that it falls off more quickly toward larger scales. This result is robust and does not depend on the value of the different hyper parameters that we set for the analysis, which we demonstrate in Section \ref{sect:validation}.

A more detailed analysis of the advantages of this dust removal is shown in Section \ref{sect:validation:cibxkappa_errors}. For the CIB results of \citet{planck2016_xvii} based on the GNILC component separation, we find a weaker cross correlation with the CMB lensing signal, presumably due to an oversubtraction of the CIB. This is in agreement with the findings in \citet[][Appendix A]{maniyar2019}.

\section{Validation}
\label{sect:validation}

The key challenge in the separation of CIB emission and Galactic dust emission is the validation: How can we convince ourselves that we removed all the Galactic dust but none of the CIB signal? How do we quantify the dust residuals?

\subsection{Internal validation}

\subsubsection{Estimating uncertainties for the power spectra}
\label{sect:validation:cl_errors}

\begin{figure}[tp]
	\includegraphics[width=\columnwidth]{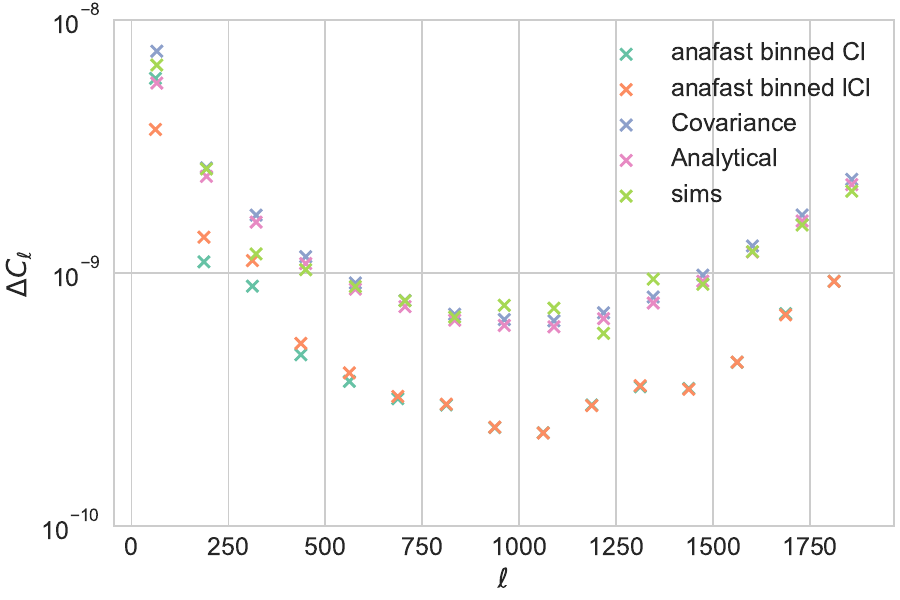}
	\caption{Comparison of different estimators for the error on the cross power spectrum of CIB and CMB lensing convergence. We find that the analytical error bars agree very well with the covariance estimate from PyMaster, as well as with the values we get from running the $C_{\ell}$ estimation many times on simulations. Simply binning the full-resolution power spectra obtained from an estimator that does not correct for the mode--mode coupling underestimates these errors significantly. For the latter point, we have used \texttt{healpy.anafast} and binned the power spectrum both in $C_{\ell}$ and in $\ell\,C_{\ell}$.}
	\label{fig:cl_errors}
\end{figure}

The computation of the error bars for the power spectra is not straightforward, as our comparison of different measures shows (Figure \ref{fig:cl_errors}). Here, we compare different estimators for the uncertainty. First, the analytical approximate estimate of the error on the CIB--CMB lensing cross power spectrum assuming a Gaussian signal and neglecting mode coupling is given by

\begin{equation}
	\left(\Delta C_{\ell}^{T\kappa}\right)^2 = \frac{\left(C_{\ell}^{T\kappa}\right)^2 + \left(C_{\ell}^{TT} + N_{\ell}^{TT}\right)\cdot \left(C_{\ell}^{\kappa\kappa} + N_{\ell}^{\kappa\kappa}\right)}{(2\ell+1)f_{\rm sky}\Delta\ell}\ . 
	\label{eq:cl_error_analytical}
\end{equation}
The $C_{\ell}$ and the $N_{\ell}$ describe the theoretical signal and noise power spectrum of the CIB temperature $T$ and the lensing convergence $\kappa$. $f_{\rm sky}$ is the effective sky fraction of the maps, and $\Delta \ell$ is the number of $\ell$-modes per bin. Note that since we use the theoretical CIB and lensing signal, we do not capture the uncertainties coming from foreground residuals this way.

We also compute the error bars based on the diagonal of the NaMaster-based covariance matrix of the power spectrum estimate, which we demonstrate to be a reliable measure and hence use throughout this work. Both the analytical and the covariance estimates agree well with the results we obtain from running 100 Gaussian sky simulations. For these, we simulate the CIB and the lensing signal and noise fields, and then recover the power spectra via \texttt{PyMaster} in the presence of the mask. The input power spectra for the CIB, the CMB lensing, and their correlation are taken from \citet{planck2014_xxx}, \citet{planck2014_xviii}, and \citet{maniyar2018}, respectively.

On the other hand, simply measuring the error bars by binning the results from an estimator that does not correct for the mode--mode coupling underestimates the error significantly.

\subsubsection{$N_{\ion{H}{i}}$ thresholds}
\label{sect:validation:nhi_thresh}

We investigated how different Galactic plane masks and $N_{\ion{H}{i}}$ thresholds affect our results. The goal is to find the optimal trade-off between a large sky fraction and a small contamination by CO-dark molecular gas and Galactic dust (cirrus).

\begin{figure*}[tp]
	\includegraphics[width=\columnwidth]{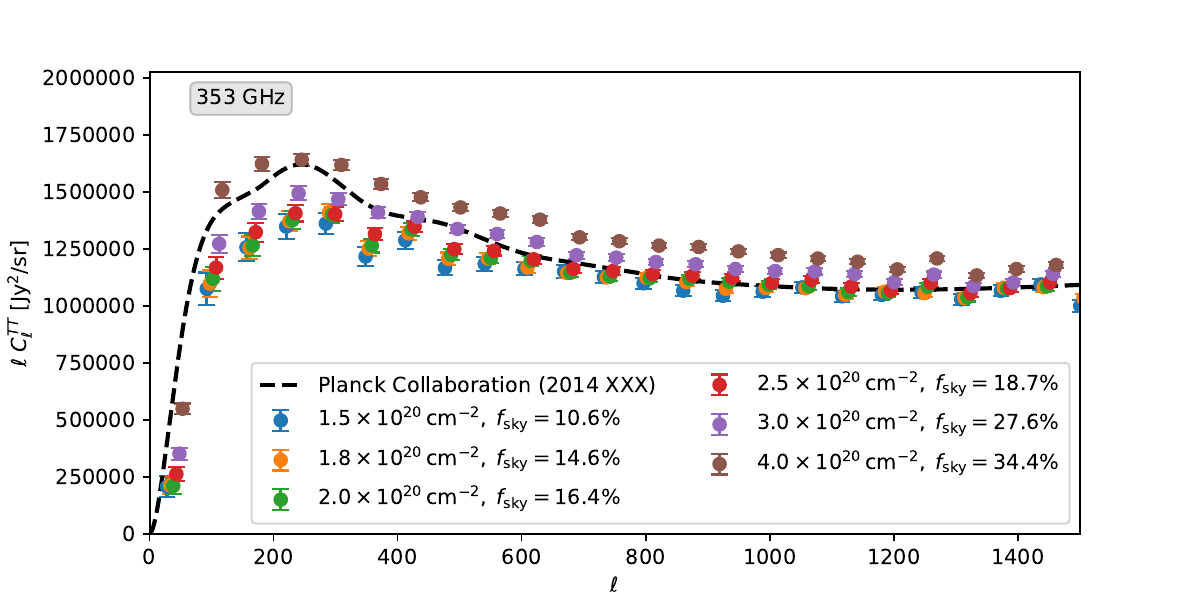}
	\includegraphics[width=\columnwidth]{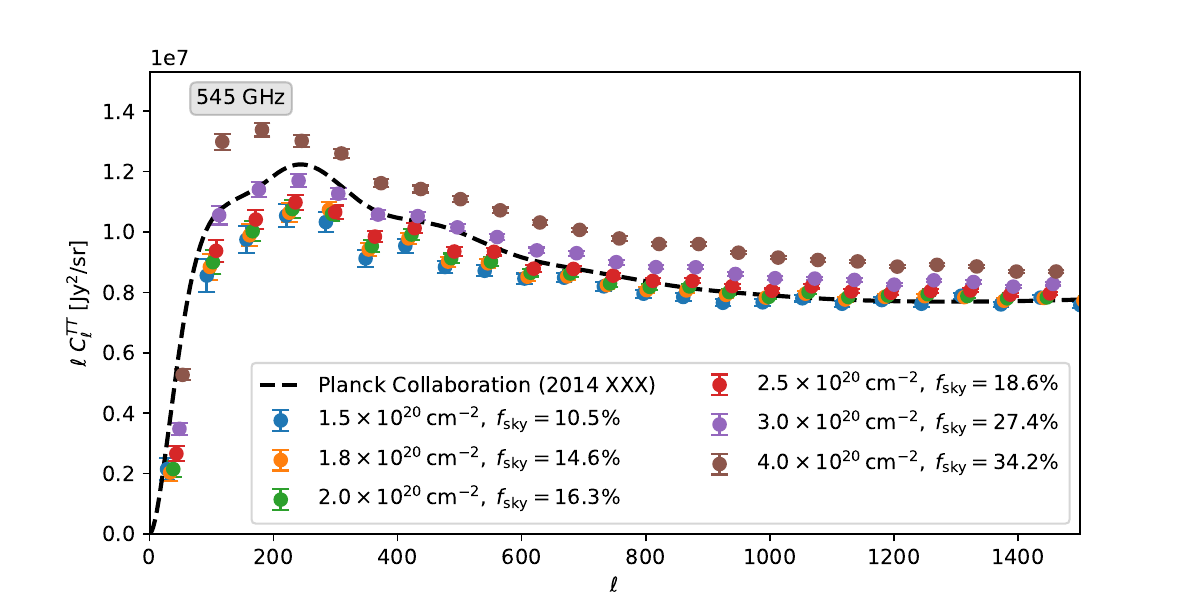}
	\includegraphics[width=\columnwidth]{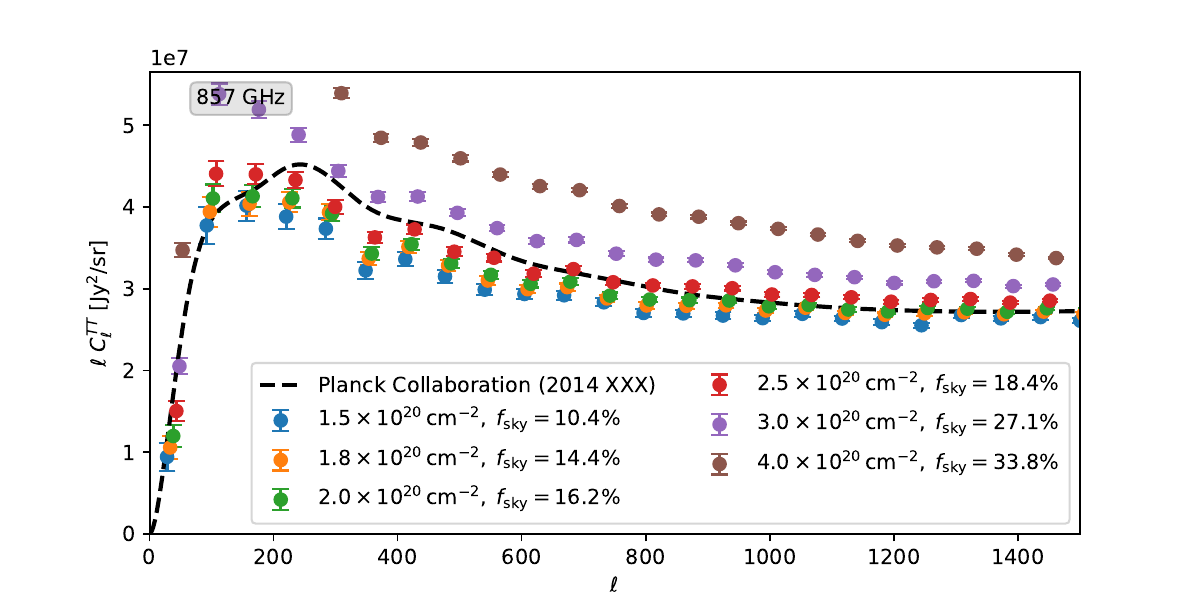}
	\caption{CIB auto power spectra for the different $N_{\ion{H}{i}}$ thresholds, highlighting the additional contamination by Galactic dust on the larger scales. The different panels correspond to the different \planck frequencies, showing the 353, 545, and 857\,GHz bands (top left, top right, bottom left). As expected, the residual dust contamination is strongest toward higher frequencies. Here, the exact choice of the $N_{\ion{H}{i}}$ threshold has the biggest impact. For reference, the dashed line shows the best-fit model from \citet{planck2014_xxx}.}
\label{fig:validation_nhi_thresholds_cibxcib}
\end{figure*}

To this end, we test different threshold values and inspect the CIB auto power spectrum (Figure \ref{fig:validation_nhi_thresholds_cibxcib}). We find that convergence is only reached for very low \ion{H}{i} column densities, which does greatly reduce the available sky fraction. For all results presented here, we decided to limit the analysis to regions with $N_{\ion{H}{I}} < 2.5\times10^{20}\rm cm^{-2}$, which is a good compromise of residual dust at small scales and a large sky fraction of \fsky\%. This is further confirmed in Section \ref{sect:validation:cibxkappa_errors}, where we study the uncertainty of the CIB--CMB lensing cross correlation in more detail.

\subsubsection{Size of the patches}
\label{sect:validation:superpix}

\begin{figure*}[tp]
	\includegraphics[width=\columnwidth]{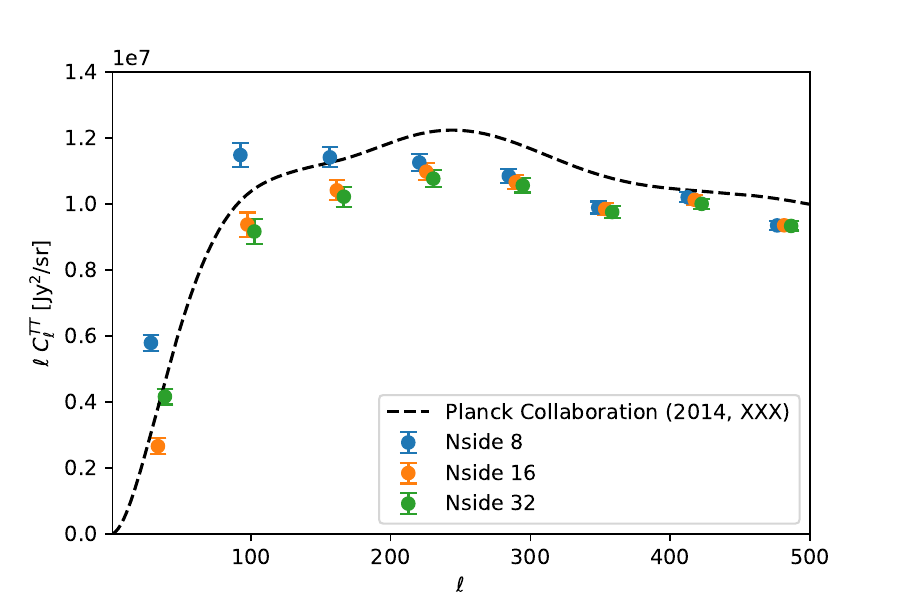}
	\includegraphics[width=\columnwidth]{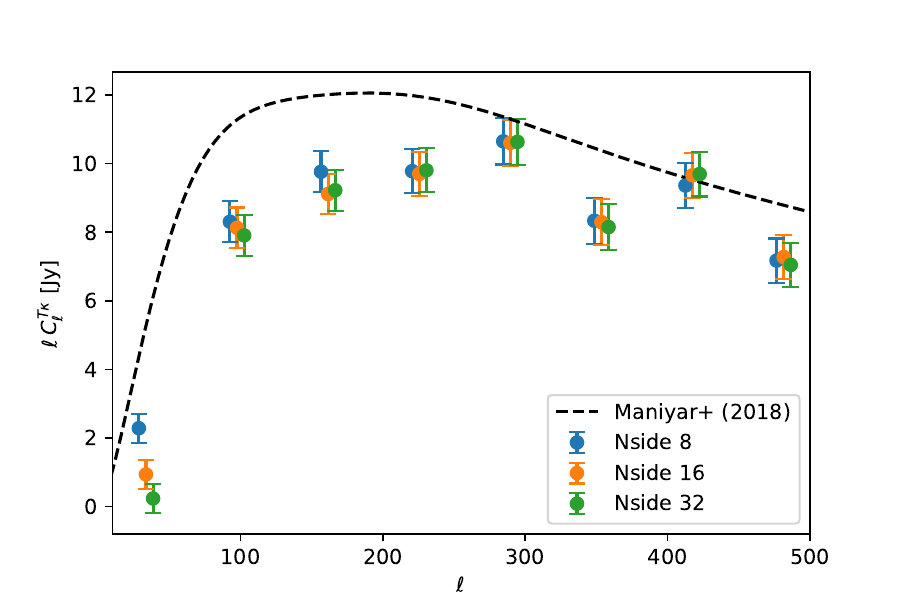}
	\caption{CIB auto power spectra and cross power spectra with the lensing potential for different sizes of the HEALPix superpixels (all at 545\;GHz). We find that working with large patches (Nside 8) leads to strong dust contamination because the spatial variations of dust-to-gas ratios cannot be captured. We also find that the CIB--CMB lensing cross correlation is unaffected by the choice of the patch size, and that $C_{\ell}$ is unbiased for $\ell \gtrsim 70$.}
\label{fig:validation_nside_superpixels}
\end{figure*}

We also investigated the different choices of the Nside of the superpixels, in which we locally perform the component separation. If these superpixels are chosen to be too large, then the complexity of the \ion{H}{i}/FIR relation will not be properly captured. If they are too small, the number of independent sight lines will be too small, and the model will eventually remove CIB structures due to chance correlations with the \ion{H}{i} (see also Section \ref{sect:spatial_structure}).

We hence repeat the CIB component separation for different choices of the superpixel sizes. Here, we use HEALPix Nsides ranging from 8 to 32. We evaluate the results for the CIB auto power spectrum and for the CIB--CMB lensing cross correlation (Figure \ref{fig:validation_nside_superpixels}). The former is difficult to interpret, because there are two effects at play. First, a smaller patch enables us to capture the Galactic dust better, reducing the contamination of the CIB at large scales. Second, the high-pass filtering of the CIB angular power spectrum due to working on individual patches could also potentially remove power on these scales (see Section \ref{sect:spatial_structure}).

To overcome this degeneracy, the CIB--CMB lensing cross power spectrum can be conveniently used because residual Galactic dust will have a very different effect. Instead of biasing the power spectrum, the dust will only lead to a higher noise, because it is uncorrelated with the CMB lensing signal.

Based on the analysis of the cross power spectrum of CIB and CMB lensing, we find that the additional noise from residual Galactic dust is very small, even for very large patches (Nside 8) for which the CIB auto power spectrum is strongly contaminated. We furthermore find that the high-pass filtering of the power spectrum due to working on individual patches is a minor effect and only affects the results below $\ell \lesssim 70$, where it biases the result by approximately $2\sigma$. Based on the present analysis, we choose an Nside of 16 for the component separation.

\subsubsection{Histograms and Gaussianity}
\label{sect:validation:hist}

We also inspected the histograms of the recovered CIB at different frequencies to see whether our reconstruction is Gaussian-distributed. In particular, we look for signs of residual CO-dark molecular gas, which would be seen as a heavy tail toward positive residual values.

\begin{figure*}[tp]
	\includegraphics[width=0.33\textwidth]{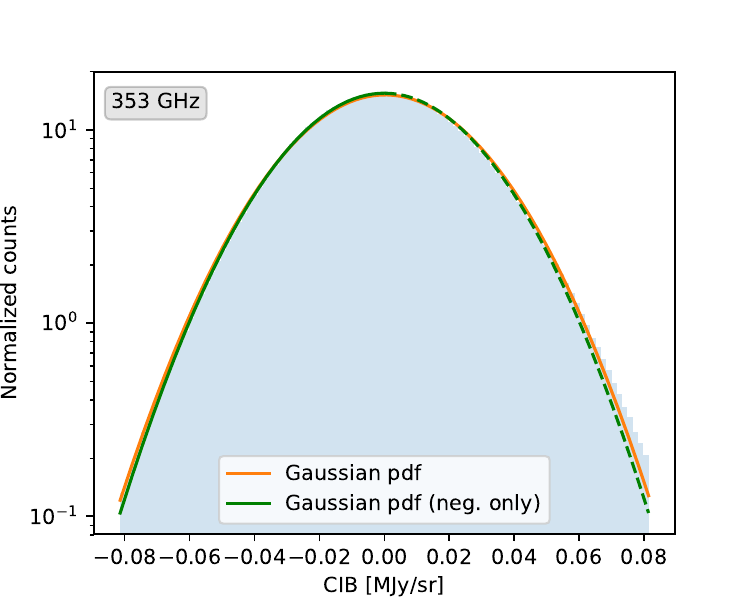}
	\includegraphics[width=0.33\textwidth]{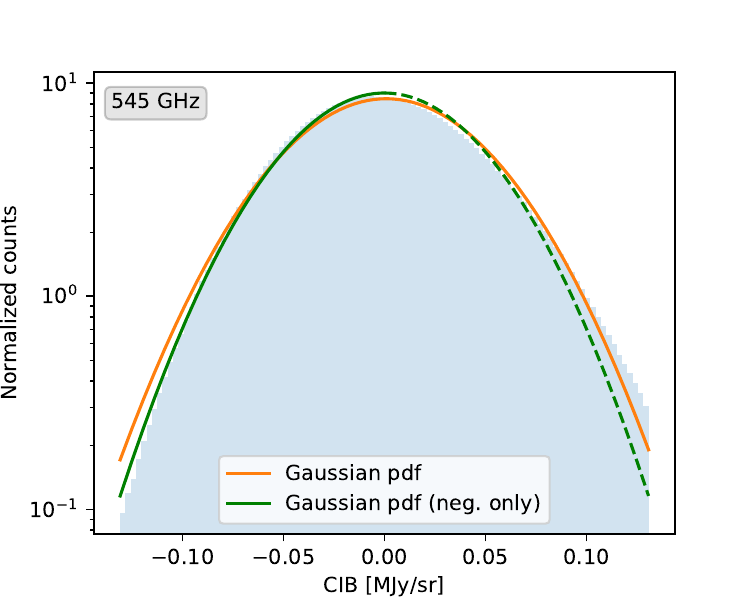}
	\includegraphics[width=0.33\textwidth]{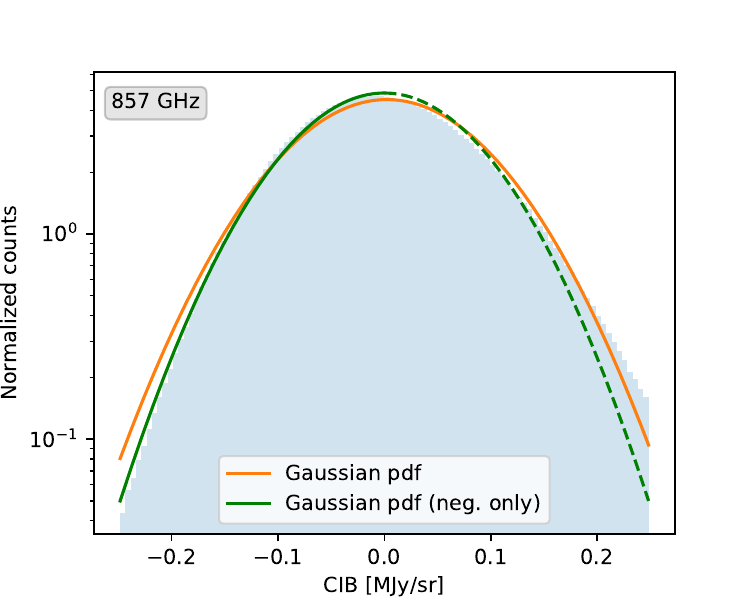}
	\caption{Histogram of CIB values for the different frequencies, including a fit to a Gaussian probability density function (pdf). We perform two different Gaussian fits the histograms. The blue line fits the entire range of the pdf, whereas the green line is only fit to the rising flank of the pdf (solid part) and then extrapolated to positive values (dashed part). This highlights how CO-dark molecular gas contaminates the maps, and how its effect is stronger at the higher frequencies.}
\label{fig:validation:histogram}
\end{figure*}

We present the histogram in Figure \ref{fig:validation:histogram}. To quantify the distributions, we fit a Gaussian to the histograms, both for the full data set and only for the rising flank that is uncontaminated by the dark gas.

As expected, we find the imprint of the dark gas, visible as an excess toward positive residuals. This scales with frequency, where the higher frequencies are more prone to contain Galactic dust instead of CIB signal. Especially for the 857\;GHz band, we observe that we are probing the emission close to the peak of the modified blackbody spectrum of Galactic dust.

These histograms are also the basis for the residual masking scheme, which iteratively removes the foreground, evaluates the histogram of the resulting CIB, and masks outliers that exceed a threshold of $3\sigma$.

\subsubsection{Offset in the FIR/\ion{H}{i} relation}
\label{sect:validation:offset}

Lastly, we also omitted the offsets in the FIR/\ion{H}{I} relation, not allowing a spatially varying zero-point. We find that this leads to widespread failure of the component separation, which is visible both in the map and in the auto power spectrum. This is consistent with the spatially varying offset that we presented in Figure \ref{fig:offsets}, and which shows a large-scale variation of this zero-point. One advantage of this zero-point is that it can capture dust emission unrelated to the \ion{H}{i}, such as that from the warm ionized medium (WIM) \citep{lagache1999}. With Galactic dust dominating the CIB signal on large scales, this additional zero-point is also well suited to address spatial variations in the FIR/\ion{H}{i} relation without biasing the CIB signal.

\subsubsection{Uncertainty of the CIB--CMB lensing cross correlation}
\label{sect:validation:cibxkappa_errors}

\begin{figure}
	\includegraphics[width=\columnwidth]{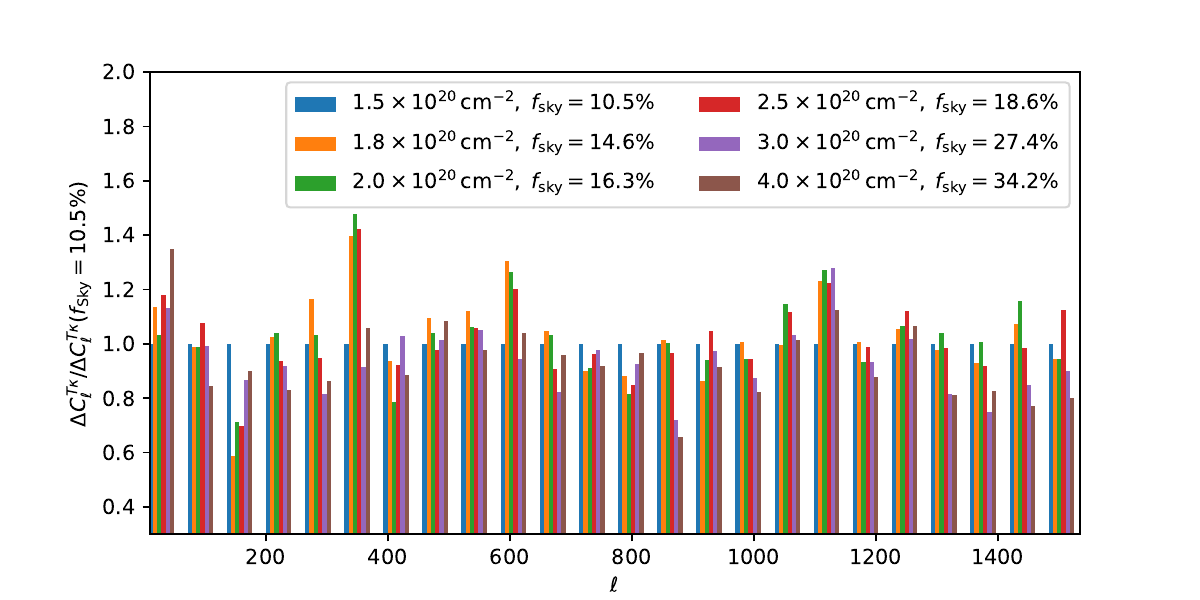}
	\caption{Errors for the CIB--CMB lensing cross correlation for different sky fractions, normalized to the smallest sky fraction. While an increased sky fraction leads to a larger sample and thus to smaller error bars, the increasing dust contamination counteracts this effect. We find that these two opposing effects cancel out almost completely, yielding constant error bars for $C_{\ell}^{T\kappa}$ independent of the sky fraction that is analyzed.}
	\label{fig:cibxkappa_errors}
\end{figure}

An important question to ask is how much of an impact the selected sky fraction has on the precision of the power spectrum. For the CIB--CMB lensing cross correlation, two effects are at play: a larger sky fraction implies a larger sample and hence smaller errors, but this additional area is also subject to dust residuals, which increase the errors. We present a comparison of the error bars for the CIB--CMB lensing correlation $C_{\ell}^{T\kappa}$ in Figure \ref{fig:cibxkappa_errors}. These error bars are computed by simply binning the full-resolution power spectra and then evaluating the standard deviation in each bin. While we do not use this for any of the final error bars in our analysis, it is useful here because it allows us to compare the relative level of the errors without having to assume the underlying theory power spectra, which is particularly difficult for the residual dust contamination.

We find that for a wide range of sky fractions ($10\%$ -- $34\%$), the two effects described above cancel out almost perfectly, and the derived uncertainties are mostly independent of the sky fraction. While this holds for the CIB--CMB lensing cross correlation, the effect is very different for the CIB auto power spectra. Here, the residual dust acts as a bias and not as an uncorrelated quantity that increases the noise (see Figure \ref{fig:validation_nhi_thresholds_cibxcib}).

\begin{figure}
	\includegraphics[width=\columnwidth]{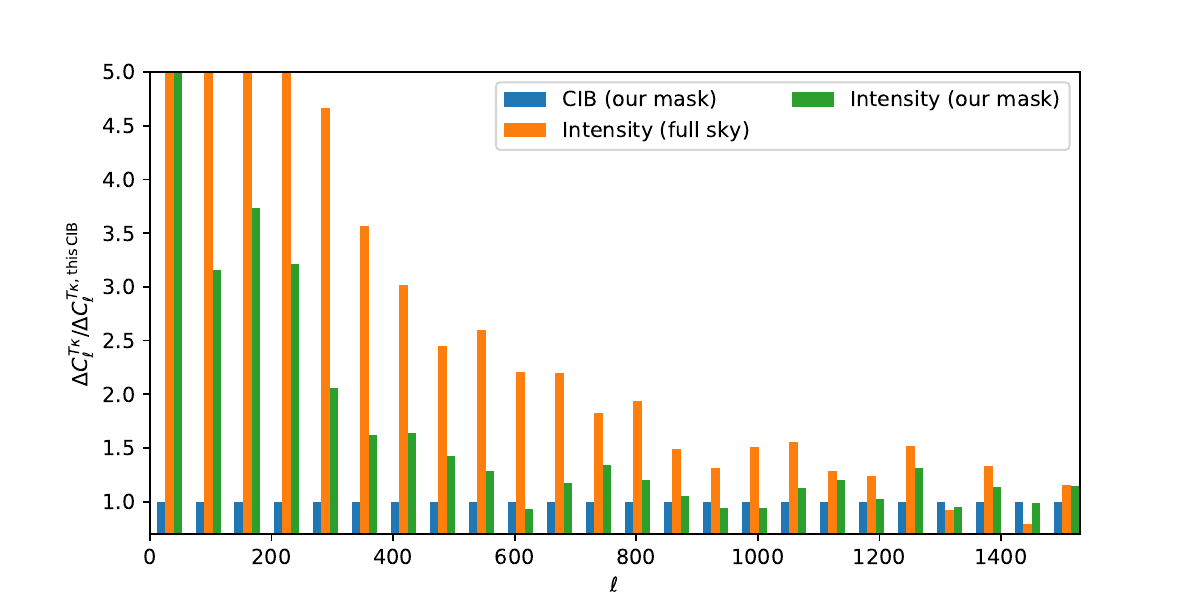}
	\caption{Errors for the cross correlation of different FIR intensity/CIB maps with the lensing convergence. We further normalize this to the errors from our best fit CIB map. The different bars show our best fit CIB map (blue), the raw FIR intensity in our footprint (green), and the raw FIR intensity for 66\% of the sky (orange, only constrained by the lensing mask). This illustrates the large impact of the foreground removal on the large scales that are dominated by the infrared cirrus. Even at small spatial scales ($\ell \sim 1500$), the residual dust contamination does not allow tight constraints on the raw intensity-based power spectra, despite the much larger sky fraction ($\fsky\%$ vs. $66.9\%$).}
	\label{fig:intensityxkappa_errors}
\end{figure}

To quantify the effect of the foreground dust removal on the CIB--CMB lensing cross correlation, we compare the error bars for this cross correlation for different CIB maps. First, we use the CIB map derived here. Second, we use the raw, uncorrected FIR intensity within the footprint presented here. Lastly, we use the FIR intensity for the entire sky area for which lensing data are available ($66.9\%$). The result (Figure \ref{fig:intensityxkappa_errors}) shows that for the large scales ($\ell \lesssim 800$), the foreground removal is crucial to optimize the accuracy of the cross correlation. Even at smaller scales, this cleaning allows our results to be equivalent to or better than the raw intensity for a sky fraction that is 3.7 times larger ($\fsky\%$ vs. $66.9\%$).

\subsection{External validation}

\subsubsection{CIB maps from Planck XXX}

Aside from the comparison of the power spectra derived here with previous analyses, we also analyzed the differences at the map level. Here, we use the 545\,GHz channel and compare our results to the those obtained in \citet[][P14]{planck2014_xxx}. We inspected the smaller individual fields which are based on data from the Green Bank telescope (GBT), as well as the larger so-called GASS field.

To perform this comparison, we convolved the higher-resolution data from P14 to the $5'$ resolution of the CIB maps presented here.

\paragraph{Small fields}

\begin{figure*}
\centering
	\subfloat[N1 field]{
		\includegraphics[width=0.49\textwidth]{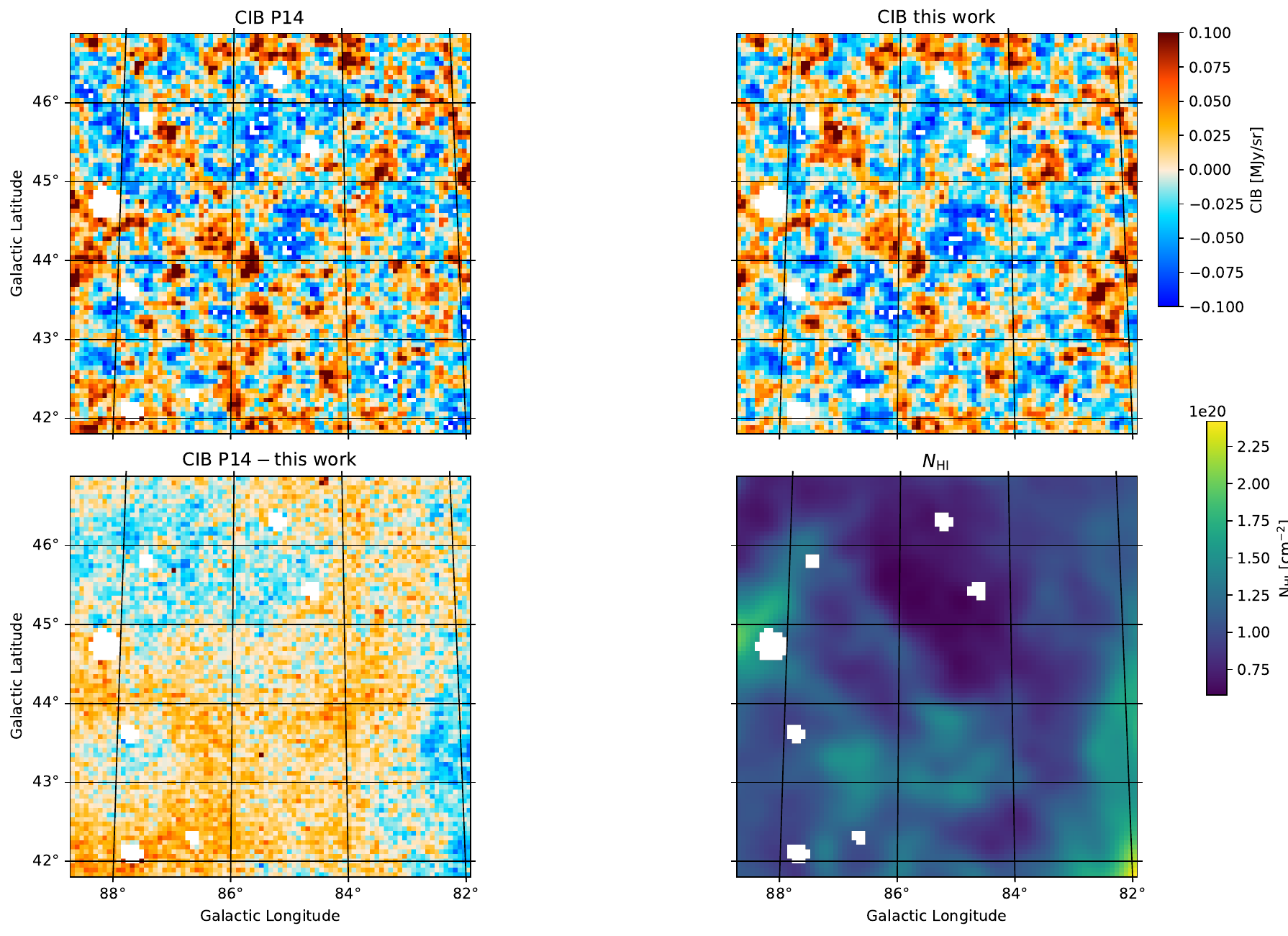}
	}
	\subfloat[AG field]{
		\includegraphics[width=0.49\textwidth]{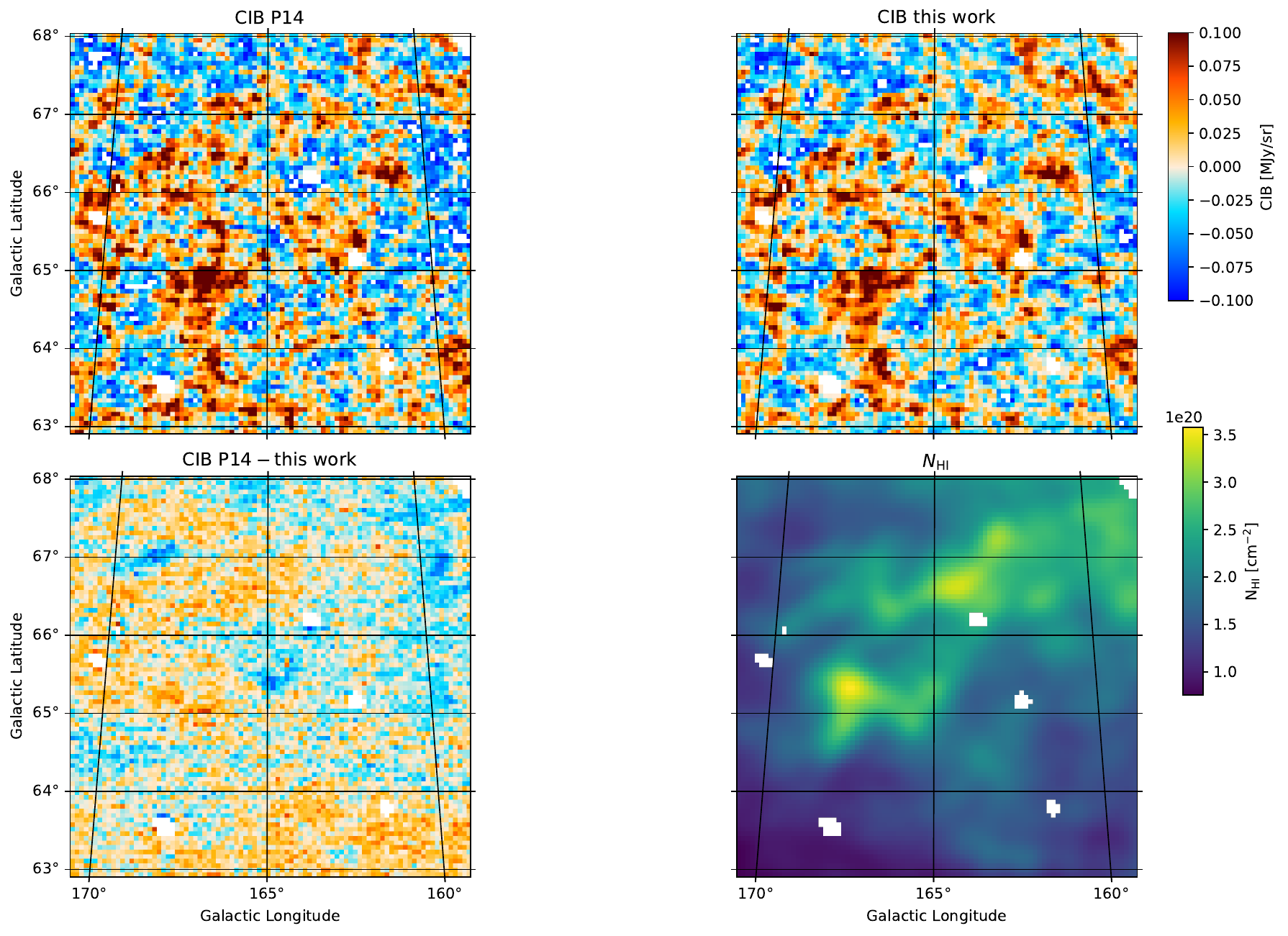}
	}
	\caption{Comparison of the CIB fields derived in the present work with those from P14. We show the N1 field (\textbf{left}) and the AG field (\textbf{right}) as examples, the other figures can be found in Appendix~\ref{sect:app:comparison}. For each field, we show the CIB images from this work (\textbf{top left}) and from P14 (\textbf{top right}), their differences (\textbf{bottom left}), and the \ion{H}{i} column density.}
	\label{fig:p14_comp_maps}
\end{figure*}

The results of this comparison are shown in Figure \ref{fig:p14_comp_maps}. We find that the two CIB estimates show great agreement at the map level, with the differences being dominated by large scales. Several factors contribute to these differences.

First, our work and that conducted in P14 differ fundamentally in the way the dust cleaning is done spatially. For the 2014 results, each field was analyzed separately, and the dust-to-gas ratios were determined for the entire field. Our work uses smaller patches, based on the HEALPix grid, and performs the dust cleaning in each of these separately. Nonetheless, we point out that no sign of this underlying structure can be seen in the CIB difference maps.

Second, the underlying \ion{H}{i} data, as well as the selection of the \ion{H}{i} data in velocity space, differ between the present work and that in P14. For the small fields that are the basis for the comparison shown here, P14 used data from the GBT that was later published as the GHIGLS data set \citep{martin2015}. In their analysis, the \ion{H}{i} data are individually binned into maps of \ion{H}{i} column density that represent the physical phases of the low- and intermediate-velocity clouds (LVC/IVC) \citep{putman2012}. In our analysis presented here, we use the slightly lower-resolution HI4PI data, and use the GLM to select features in velocity space, without relying on a manual selection (Section \ref{sect:glm_method}).

Lastly, additional higher-order effects include different calibrations for the \planck data (PR1 vs. PR3), and slightly different resolutions due to different reprojection strategies.

\paragraph{The GASS field}

\begin{figure*}
	\centering
	\includegraphics[width=\textwidth]{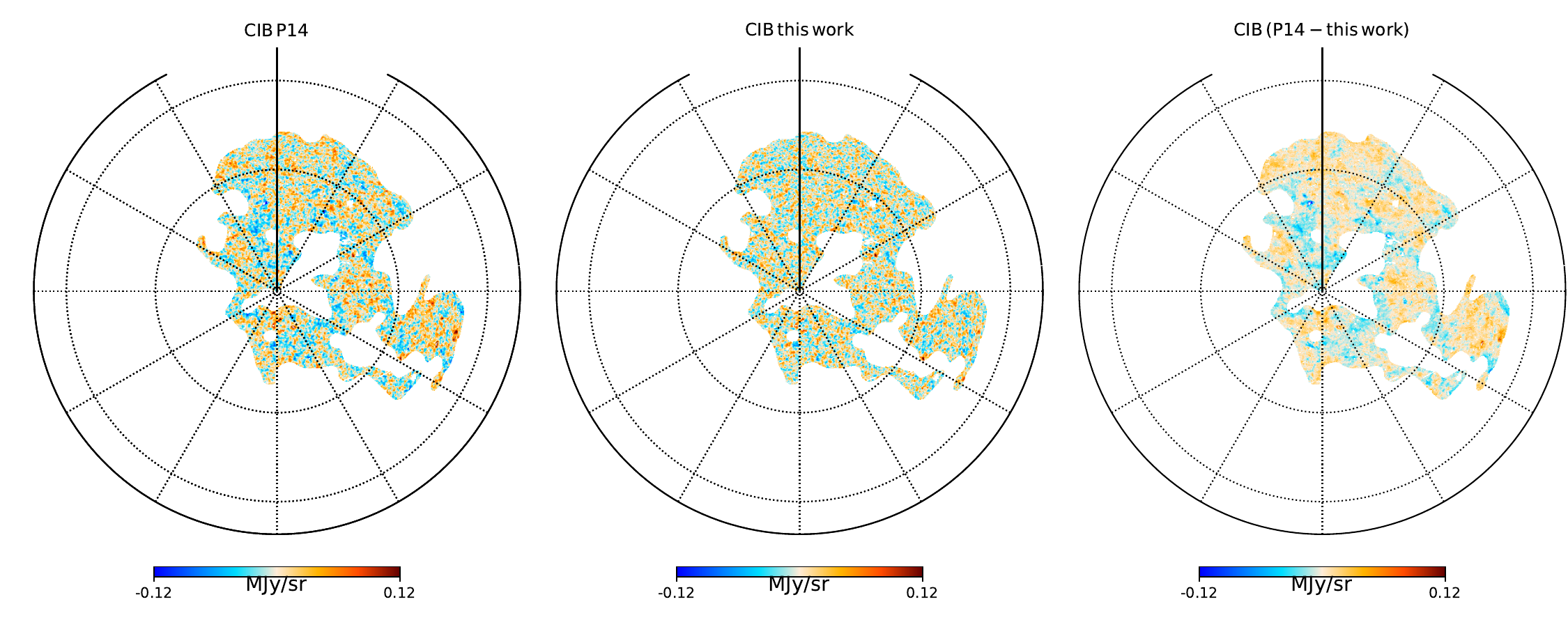}
	\caption{Comparison of the 545\,GHz CIB images of the southern Galactic cap. \textbf{Left:} P14 data. \textbf{Center:} the present work. \textbf{Right:} difference between the two.}
	\label{fig:p14_comp_gass_maps}
\end{figure*}

We conduct the same comparison for the larger GASS field from P14 (Figure \ref{fig:p14_comp_gass_maps}). Here, we bring both CIB maps to the GASS resolution of $16.2'$. While our analysis of this field does not differ from the one in the previous section, the P14 approach differs from that applied to the smaller fields.

For the smaller fields, it was assumed that the dust-to-gas ratio is constant within the field, and only varies for the HVC/IVC/LVC phases. Moreover, the resolution of the GBT \ion{H}{i} data is $9'$, which is close to the \planck resolution at these frequencies ($4'$--$5'$). The larger GASS field in contrast is too large to assume a constant dust-to-gas ratio, hence a map of dust emissivities and offsets is constructed for patches with a diameter of $15^{\circ}$, centered on the HEALPix pixels with Nside 32. These parameter maps are then smoothed, very similar to the analysis presented here. The main difference from the present study is that P14 only used the map of total local \ion{H}{i} column density map for the GASS field, while we use the velocity-resolved data and apply the GLM. This additional degree of freedom now allows us to account for different emissivities along the line of sight, thereby reducing the residual dust contamination.

\subsubsection{CIB maps from GNILC}

A different approach to extract CIB maps from \textit{Planck} data has been presented in \citet{planck2016_xlviii}. Instead of using \ion{H}{i}-based dust templates, the authors implement the GNILC component separation technique. This is based on the fundamentally different angular power spectra of Galactic dust and the CIB. These maps cover over 60\% of sky and are available for the 353, 545, and 857\;GHz bands. It should be noted that the primary goal of the GNILC maps is to clean Galactic dust from CIB emission to better determine the properties of Galactic dust such as temperature $T$ and dust spectral index $\beta$. 

The most straightforward comparison can be obtained through cross-correlating the CIB against the CMB lensing convergence, which we present in Figure \ref{fig:cl_cibxkappa}. There is no straightforward way to bias this cross correlation toward a higher correlation, but it can easily be lowered by not capturing the full CIB emission. This is the case for the GNILC maps, which were designed to clean the Galactic dust from CIB contamination. The cross correlation with the CMB lensing is systematically lower than what is found in \citet{planck2014_xxx}, as already shown in \cite{maniyar2019}, and also in this work.

\section{Conclusions}
\label{sect:conclusions}


We have presented new large-scale $5'$ CIB maps, based on the \planck data and an \ion{H}{i}-based foreground removal strategy. These maps are made publicly available for the 353, 545, and 857\;GHz data, and for a range of sky fractions.

The foreground removal strategy has been advanced to be more robust, automatic, and to include more information on the three-dimensional \ion{H}{i} data. We have validated our results through comparisons with previous studies and find strong agreements both for the images and for the angular power spectra. Additional internal validation shows that the Galactic dust residuals are present down to the smallest scales, depending on the frequency and on the \ion{H}{i} threshold that is applied to the maps. 

Consequently, different maps should be used for different purposes: studies that are based on the CIB auto power rely on the smallest dust contamination possible, and should therefore use the more aggressive masks. More specifically, we recommend using a thresholds of $2.5\times 10^{20}\;\rm cm^{-2}$ (353 GHz), $2.0\times 10^{20}\;\rm cm^{-2}$ (545 GHz), and $1.8\times 10^{20}\;\rm cm^{-2}$ (857 GHz). For cross correlation analyses, these dust residuals are only a source of noise and do not introduce a bias, hence a larger sky fraction might be preferable to increase the sample size.

For the CIB--CMB lensing cross correlation, we find that the peak is at smaller scales than presented in previous studies, and that the power spectrum falls off slightly faster toward larger scales. This is independent of the various parameters such as sky fraction and patch size for the dust modeling.

Despite the excellent agreement with previous studies, we note that the calibration differences between the three \planck releases complicate the comparison, and a complete cross calibration analysis between the different releases would be required. Part of the discrepancy could come from different absolute brightnesses in the different releases \citep[up to 3.4\% at 857\,GHz, ][]{planck2015_viii}.

We anticipate that these maps will be particularly useful for cross-correlation studies with other tracers of the large-scale structure, such as the CMB lensing signal and galaxy surveys.


We make all our CIB data products available online \footnote{\url{https://doi.org/10.7910/DVN/8A1SR3}}. This includes the following:

\begin{enumerate}
	\item The CIB maps at 353, 545, and 857\;GHz. We provide these for the \planck full mission data, as well as for the odd-/even-ring data splits.
	\item The \ion{H}{i}-based dust model for all frequencies.
	\item Each map comes with a mask, where we proved both the boolean and the apodized masks.
	\item The effective window functions and the FWHM of the maps, which are required to deconvolve the angular power spectra.
	\item Binned power spectra for the different CIB cross correlation spectra $C_{\ell}^{\nu_1\nu_2}$, and for the CIB--CMB lensing cross correlation $C_{\ell}^{T\kappa}$
	\item Python notebooks and scripts that illustrate how to work with the data are available online\footnote{\url{https://github.com/DanielLenz/PlanckCIB}}.
\end{enumerate}

\subsection{Outlook}

A central difficulty in developing component separation for the Galactic dust and the CIB is the validation of the results. Simulations would be a very powerful tool to have in this context, but would require an in-depth understanding of the interplay of the gaseous and dusty ISM for the full sky and with resolutions down to $1'$.

The next step in advancing this work is to include multi-scale and multi-frequency information. The former has already been demonstrated in \citet{planck2016_xlviii}, and a combination with the work presented here would be very promising. While the frequency information would not contribute much to disentangle the two components, a combination of the \ion{H}{i}-based dust model with existing algorithms for CMB component separation could be a major advancement.

\acknowledgments
\subsection*{Acknowledgements}

Part of the research described in this paper was carried out at the Jet Propulsion Laboratory, California Institute of Technology, under a contract with the National Aeronautics and Space Administration. We acknowledge financial support from ``Programme National de Cosmologie et Galaxies'' (PNCG) of CNRS/INSU, France. D.L. acknowledges hospitality from the Laboratoire d'Astrophysique de Marseille, where part of this work was completed. Part of this work has been carried out thanks to the support of the OCEVU Labex (ANR-11-LABX-0060) and the A*MIDEX project (ANR-11-IDEX-0001-02) funded by the ``Investissements d'Avenir'' French government program managed by the ANR. This project has received funding from the European Research Council (ERC) under the European Union's Horizon 2020 research and innovation program (grant agreement No. 788212).

We are very grateful to Paolo Serra, Brandon Hensley, Abhishek Maniyar, Bryan Steinbach, Agn\`es Ferte, Tzu-Ching Chang, Fran\c{c}ois Boulanger, Eric Huff, Emmanuel Schaan, Simone Ferraro, and Matthieu B\'ethermin for long and very fruitful discussions. We thank Peter Kalberla for providing the Milky Way model of Galactic rotation.

This research has made use of NASA's Astrophysics Data System, matplotlib \citep{hunter2007}, SciPy \citep{jones2001}, NumPy \citep{vanDerWalt2011}, scikit-learn \citep{pedregosa2011}, as well as Astropy, a community-developed core Python package for Astronomy \citep{astropy2013, astropy2018}. Some of the results in this paper have been derived using the HEALPix \citep{gorski2005} package

Part of this work was carried out under the program Milky-Way-\textit{Gaia} of the PSI2 project funded by the IDEX Paris-Saclay, ANR-11-IDEX-0003-02.


\bibliography{references}

\appendix

\section{CIB maps}
\label{sect:app:maps}

\begin{figure*}[tp]
	\includegraphics[bb=0 0 600 400, clip=, width=0.8\textwidth]{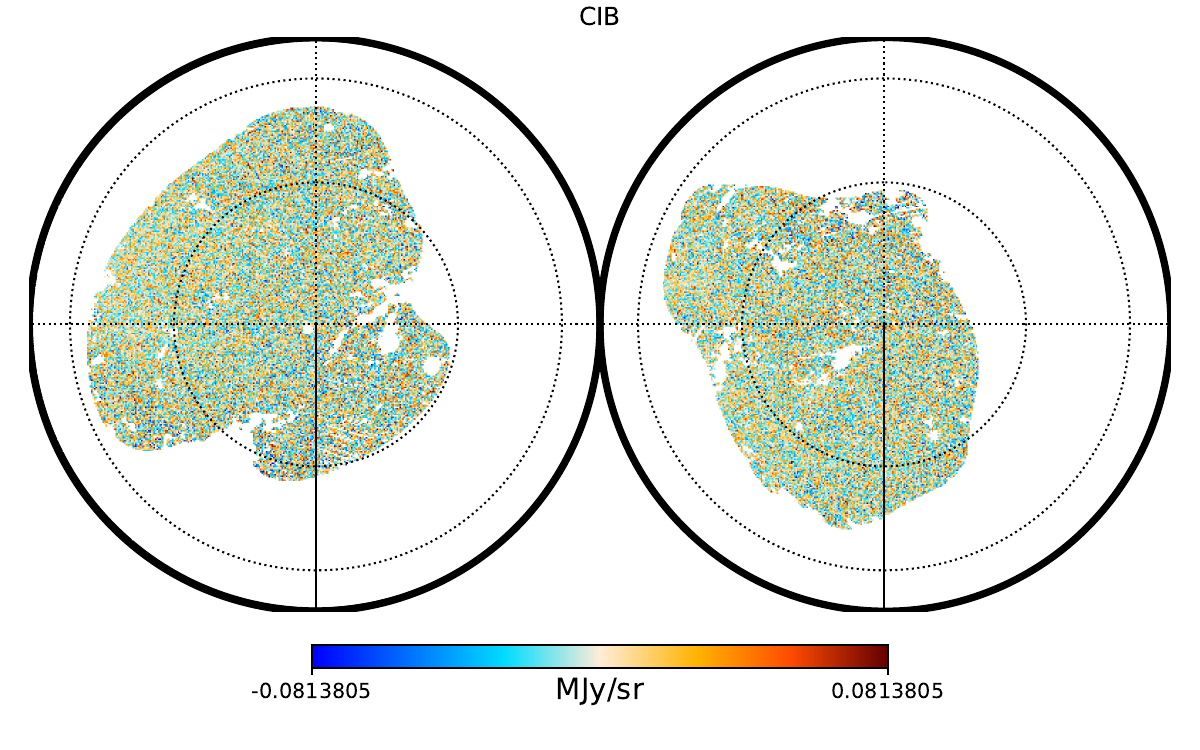}
	\caption{CIB anisotropies at 353\,GHz for the Galactic poles (north left, south right).}
\label{fig:orth_cib_353}
\end{figure*}

\begin{figure*}[tp]
	\includegraphics[bb=0 0 600 400, clip=, width=0.8\textwidth]{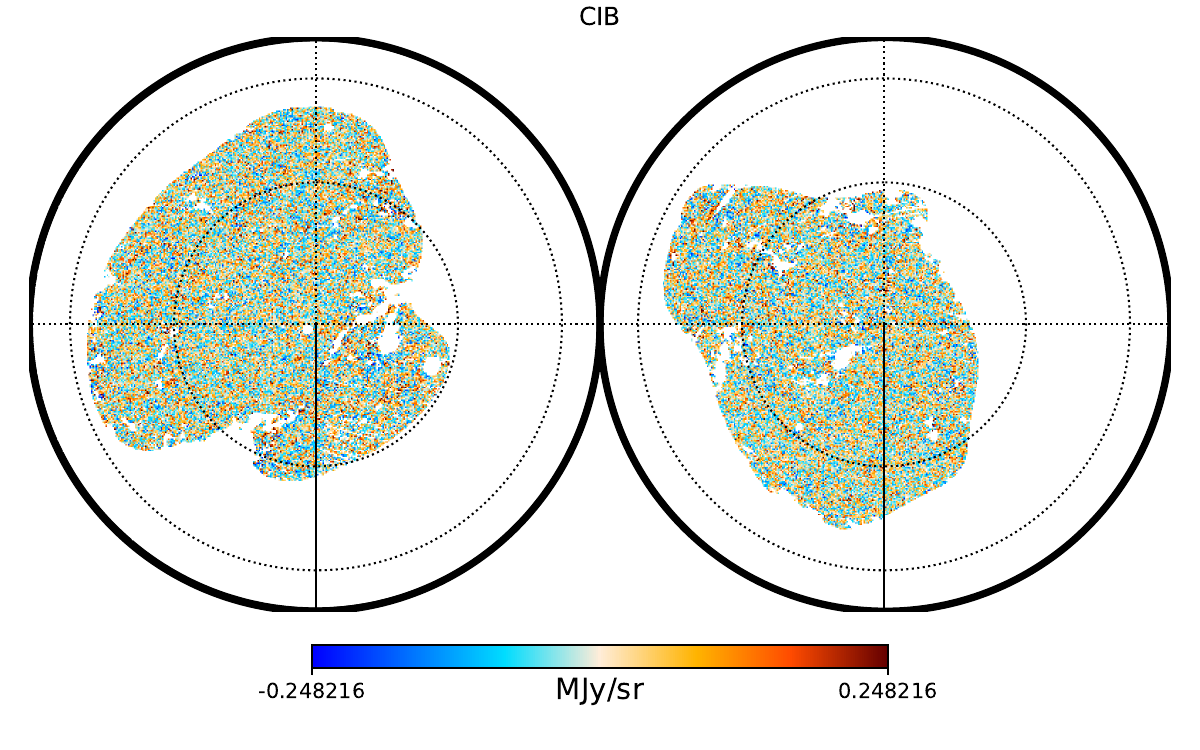}
	\caption{CIB anisotropies at 857\,GHz for the Galactic poles (north left, south right).}
\label{fig:orth_cib_857}
\end{figure*}

We present the CIB maps at 353 and 857\,GHz (Figs.~\ref{fig:orth_cib_353} and \ref{fig:orth_cib_857}). The 545\,GHz is shown in Figure \ref{fig:orth_cib}. These images are generated for an $N_{\ion{H}{i}}$ threshold of $2.5\times 10^{20}\;\rm cm^{-2}$; further maps can be obtained through the released data.

\section{Comparison with P14 results}
\label{sect:app:comparison}

\begin{figure*}
\centering
	\subfloat[SP field]{
		\includegraphics[width=0.49\textwidth]{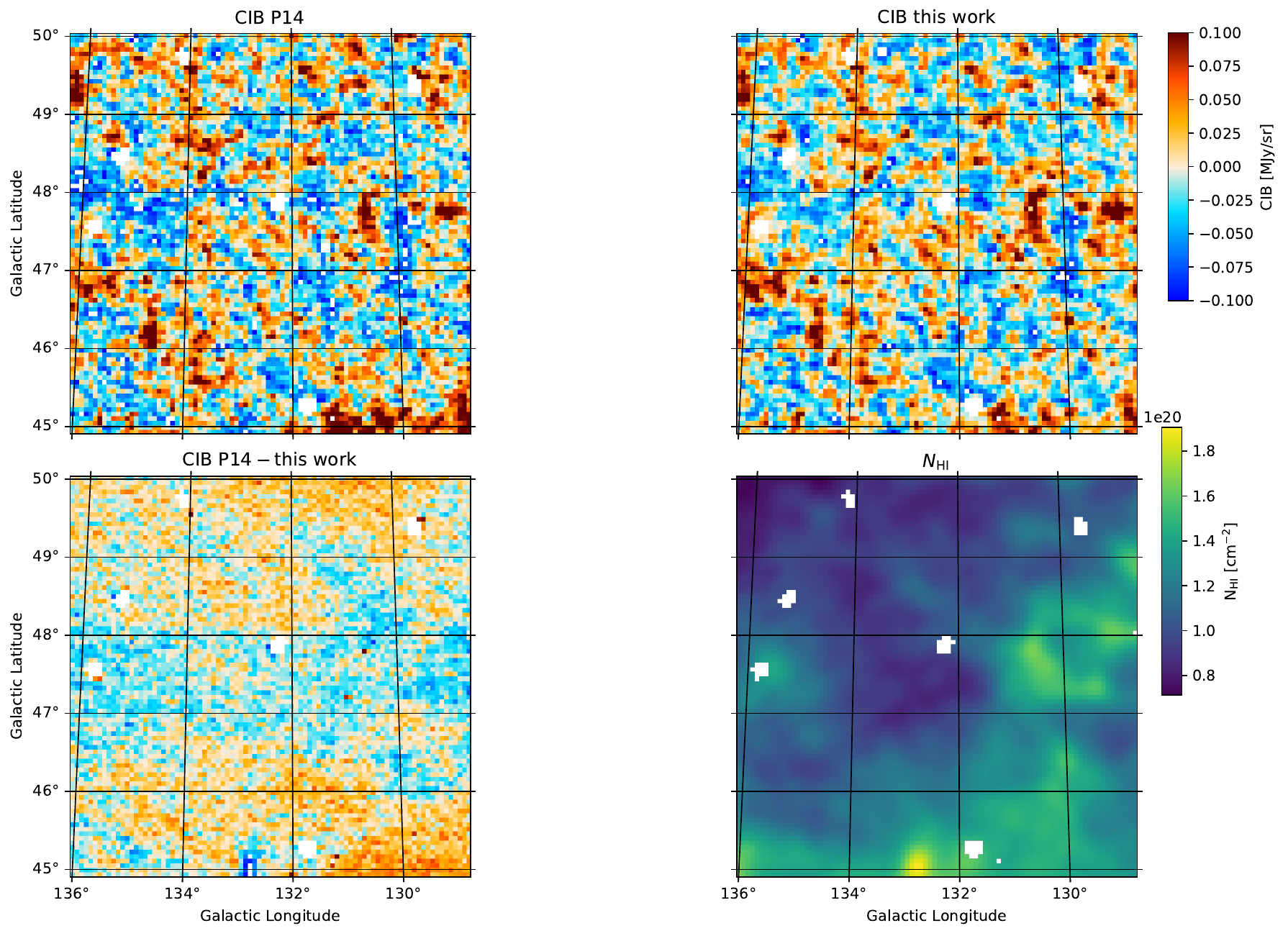}
	}
	\subfloat[Bootes field]{
		\includegraphics[width=0.49\textwidth]{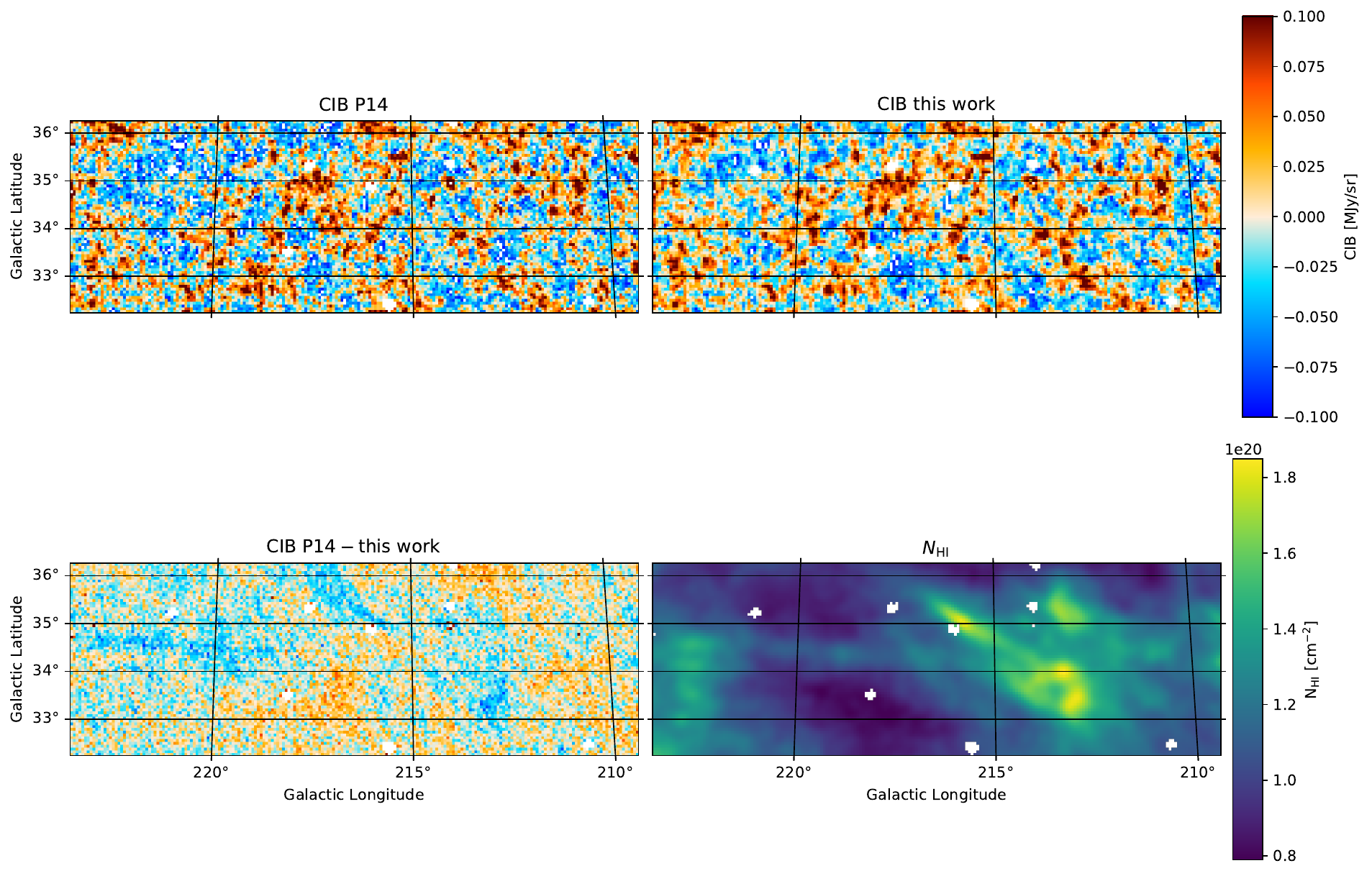}
	}
	\caption{Comparison of the CIB fields derived in the present work with the ones from \citet{planck2014_xxx}. We show the SP field (\textbf{left}) and the Bootes field (\textbf{right})}
	\label{fig:sp_bootes}
\end{figure*}

\begin{figure*}
\centering
	\subfloat[EBHIS field]{
		\includegraphics[width=0.49\textwidth]{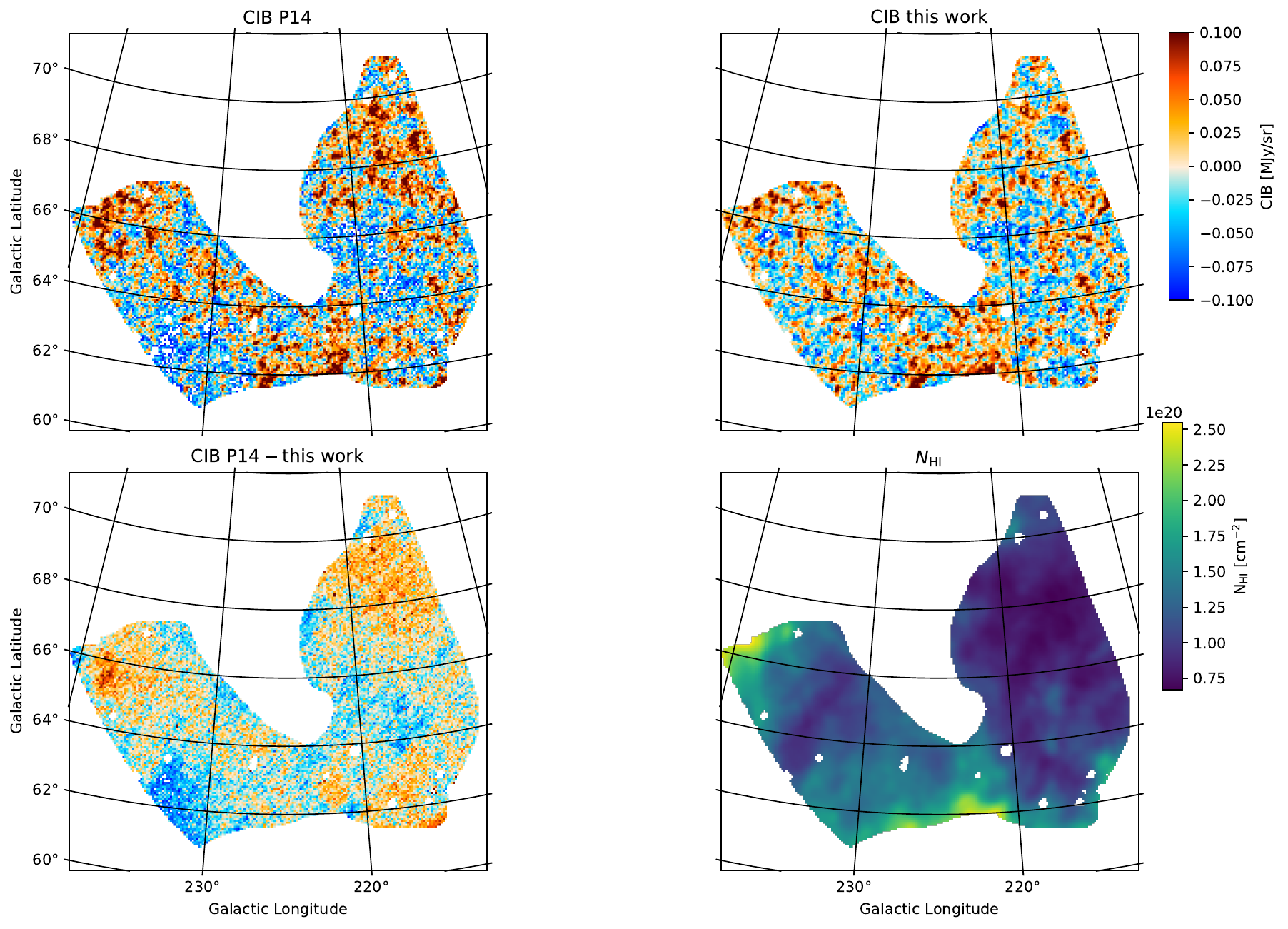}
	}
	\caption{Comparison of the EBHIS CIB field derived in the present work with the one from \citet{planck2014_xxx}.}
	\label{fig:ebhis}
\end{figure*}

We present additional map-based comparisons of the 545\,GHz CIB maps derived here with the ones presented in \citet{planck2014_xxx}. Figure \ref{fig:sp_bootes} shows the SP and the Bootes field, Figure \ref{fig:ebhis} the EBHIS field.

\end{document}